\documentclass[11pt,preprint]{aastex}

\newcommand{\simgt}{\lower.5ex\hbox{$\;\buildrel>\over\sim\;$}}
\newcommand{\simlt}{\lower.5ex\hbox{$\;\buildrel<\over\sim\;$}}

\newcommand{\msun}{\ensuremath{~M_\odot}}
\newcommand{\lsun}{\ensuremath{~L_\odot}}
\newcommand{\hi}{\rm H{\sc i}}
\newcommand{\hii}{\rm H{\sc ii}}
\newcommand{\heii}{\rm He{\sc ii}}

\newcommand{\qlyc}{Q$_{\rm Lyc}$}

\newcommand{\av}{A$_{\rm V}$}
\newcommand{\neii}{\rm [Ne\,{\sc ii}]}
\newcommand{\neiii}{\rm [Ne\,{\sc iii}]}
\newcommand{\arii}{\rm [Ar\,{\sc ii}]}
\newcommand{\ariii}{\rm [Ar\,{\sc iii}]}

\newcommand{\siii}{\rm [S\,{\sc iii}]}
\newcommand{\sii}{\rm [S\,{\sc ii}]}
\newcommand{\SiII}{\rm [Si\,{\sc ii}]}
\newcommand{\siv}{\rm [S\,{\sc iv}]}
\newcommand{\oiv}{\rm [O\,{\sc iv}]}
\newcommand{\fev}{\rm [Fe\,{\sc v}]}
\newcommand{\feiii}{\rm [Fe\,{\sc iii}]}

\newcommand{\hst}{{\sl HST}}

\newcommand{\brg}{\ensuremath{Br\gamma}}
\newcommand{\ha}{\ensuremath{H\alpha}}
\newcommand{\hb}{\ensuremath{H\beta}}

\newcommand{\mic}{$\mu${\rm m}}

\newcommand{\sbs}{SBS\,0335$-$052\,E}

\newcommand{\hh}{Haro\,3}
\newcommand{\spitzer}{{\it Spitzer}}
\newcommand{\iso}{{\it ISO}}
\newcommand{\iras}{{\it IRAS}}
\newcommand{\mopex}{{\it MOPEX}}
\newcommand{\spice}{{\it SPICE}}
\newcommand{\dusty}{{\it DUSTY}}

\shorttitle{The \spitzer\ View of \hh}
\shortauthors{Hunt et al.}

\begin{document}

\title{The \spitzer\ View of Low-Metallicity Star Formation: I. \hh}

\author{
	Leslie~K.~Hunt\altaffilmark{1},
	Trinh~X.~Thuan\altaffilmark{2}, 
	Marc~Sauvage\altaffilmark{3}, and
	Yuri~I.~Izotov\altaffilmark{4}
}
\altaffiltext{1}{INAF-Istituto di Radioastronomia-Sez.\ Firenze, 
L.go Fermi 5, I-50125 Firenze, Italy; hunt@arcetri.astro.it.}
\altaffiltext{2}{Astronomy Department, University of Virginia, P.O. Box 3818,
University Station,  
Charlottesville, VA 22903, USA: txt@virginia.edu}
\altaffiltext{3}{CEA/DSM/DAPNIA/Service d'Astrophysique, UMR AIM, CE Saclay, 
91191 Gif sur Yvette Cedex, France: msauvage@cea.fr}
\altaffiltext{4}{Main Astronomical Observatory, National Academy of Sciences of Ukraine, 
03680 Kiev, Ukraine; izotov@mao.kiev.ua}

\begin{abstract}
We present \spitzer\ observations of the
blue compact dwarf galaxy (BCD) \hh, with an oxygen abundance
of 12$+$log(O/H)\,=\,8.32. 
These data are part of a larger study of
star formation and dust in low-metallicity environments.
The IRS spectrum of \hh\ shows strong narrow 
Polycyclic Aromatic Hydrocarbon (PAH) emission, with high equivalent widths. 
Gaseous nebular fine-structure lines are also seen.
Despite the absence of optical high-excitation lines,
a faint high-ionization \oiv\ line at 25.89\,\micron\ indicates 
the presence of radiation as hard as 54.9\,eV.
A CLOUDY model suggests that 
the MIR lines originate in two regions: a low-extinction optically-emitting region,
and an optically invisible one with much higher extinction.
The morphology of \hh\ changes with wavelength.
IRAC 4.5\,\micron\ traces extended stellar photospheric emission from the 
body of the galaxy and hot dust continuum coming mainly from 
star-forming regions; 8\,\micron\ probes 
extended PAH emission coming mainly from the general ISM; 
MIPS 24 and 70\,\micron\ images map
compact small-grain warm dust emission associated with 
active star formation, and 160\,\micron\ reflects cooler extended dust 
associated with older stellar populations.
We have derived the optical-to-radio spectral energy distribution (SED) of the brightest 
star-forming region A in \hh.
The best-fit \dusty\ model of the SED 
gives a total luminosity of 2.8$\times$10$^9$ L$_\sun$
and a mass of 2.8$\times$10$^6$ M$_\sun$ for the ionizing clusters.
We infer an extinction \av $\la$3, intermediate between the optical 
\av $\sim$ 0.5 and the radio \av $\sim$ 8, 
consistent with the picture that longer wavelength observations 
probe more deeply into star-forming regions. 
\end{abstract}
\keywords{Galaxies: individual: Haro\,3; Galaxies: compact;  Galaxies: dwarf
Galaxies: starburst;  (ISM:) dust, extinction;  ISM: lines and bands;  Infrared: galaxies
\object{Haro 3}
}

\section{\label{sec:intro}Introduction}

\iso, together with COBE and SCUBA, has shown convincingly that dusty star 
formation plays a key role in the early universe. ISOCAM has identified 
a significant population of dusty 
active starburst galaxies, the ultraluminous 
infrared (IR) galaxies (ULIRG), at $z< 1.5$, which account respectively for $\simgt$30\% 
and $\sim$ 10\% of the far-IR (FIR) and submm 
backgrounds \citep{genzel00},
which in turn account for half of the total extragalactic background.
Submm observations \citep[e.g., SCUBA,][]{hughes98} suggest that the 
remaining fractions are also due to
ULIRGs, but at higher redshifts ($z>2$). 
That dust plays such an important role in the high-redshift universe may 
appear surprising, as it is usually thought that in the early 
stages of galaxy formation, when the gas is still relatively metal-free, 
dust was virtually absent.
This assumption however has been challenged by mid-IR (MIR) 
observations of one of the
most metal-poor star-forming galaxies known, \sbs\  
\citep{thuan99,plante02,houck04}.
In this local blue compact dwarf galaxy (BCD) at 
a distance of 54 Mpc and with a nebular heavy element abundance of only 
12$+$log(O/H)\,=\,7.3,
a large amount of dust (\simlt\,$10^5\,M_\odot$)
hides most ($\sim$ 75\%) of the current star-forming activity from view 
\citep{hunt01,plante02,houck04}.

Thus, to interpret the spectra of 
high-redshift galaxies, it is crucial to understand how and when this 
ubiquitous dust component is formed and how it affects the spectral energy 
distribution (SED) of galaxies 
by redistributing the ultraviolet-optical energy radiated by the 
young stars in the star-forming regions to MIR and FIR wavelengths. Such 
an understanding is needed if we are to use these SEDs to derive photometric 
redshifts and the star formation rate as a function of cosmic epoch.
Indeed,
the optical-radio SEDs of seven nearby BCDs show several significant deviations 
from the standard templates of the evolved and massive starbursts
(e.g., M\,82 and Arp\,220) 
frequently used to derive photometric redshifts of submm/mm sources
\citep{hunt05}. 
First, the location of the 
infrared bump implies generally warmer dust temperatures, 
with \sbs\ being the most extreme case of short-wavelength turnover \citep{houck04,hunt05}.
Second, the mid-IR spectra of BCDs show very weak or no Aromatic Features
in Emission (e.g., PAHs), in contrast to the strong PAH features 
seen in the spectra of  normal starburst galaxies \citep[see also][]{wu06}. 
There is evidence that these characteristics may be more common at high redshift;
the SEDs of distant SCUBA/VLA sources tend to be better fit with the 
SEDs of highly extinguished star clusters
(as in the BCD II\,Zw\,40) than with those of M\,82 and Arp\,220 
\citep{hunt05b}. 

Here we present data from our \spitzer\ Cycle 1 program 
which include spectroscopic, photometric and imaging 
observations of more than 20 BCDs with metallicities ranging from 1/30 to 1/4 
that of the Sun\footnote{Adopting the solar calibration of
\citet{anders89}.}. Besides studying the SEDs of BCDs in the infrared,
we also wish to investigate star
formation in metal-poor environments and how star formation parameters 
and dust properties change as a function of metallicity.
While much 
progress has been made in finding large populations of galaxies at high 
redshifts \citep[$z\ga\,3$][]{steidel03}, 
truly young galaxies in the process of forming remain 
elusive in the distant universe. 
The spectra of those far-away galaxies generally indicate the presence of a 
substantial amount of heavy elements, 
indicating previous star formation and metal enrichment. 
In the hierarchical picture of galaxy formation, 
large galaxies are formed from the assembly of small ones. 
Therefore,  
because of their low mass and chemically unenriched interstellar medium (ISM),
BCDs are possibly the closest examples we can find of the elementary units 
from which galaxies formed. 
Their relative proximity allows studies of their dust content with a sensitivity, 
spectral and spatial resolution that faint distant high-redshift galaxies do not permit.

In this first paper of a series, we discuss the most metal-rich object in our sample, 
\hh\ = NGC\,3353 = Mrk 35.  
\hh\ has a heliocentric radial velocity of 969 km s$^{-1}$  
\citep{thuanizotov05} which gives a   
distance of 16.3 Mpc, adopting a Hubble constant of 70 km s$^{-1}$ Mpc$^{-1}$
and correcting for the Virgocentric flow. The present distance 
is about 25\% larger than the old distance of 13.1 Mpc used by previous 
investigators who did not correct for Virgocentric flow. 
The blue apparent magnitude is of \hh\ is 13.2 (RC3), 
which corresponds to a blue absolute magnitude of $-17.9$. 

With its known population
of Wolf-Rayet stars \citep{steel96,thuanizotov05},
\hh\ is an ideal object for studying star formation in its relatively early 
stages.
Optical $V$ band observations by \citet{steel96} show that 
\hh\ hosts several star-forming regions. In order of 
decreasing brightness, there is first  
an off-center very blue star-forming 
region (A: we adopt here the nomenclature of \citealt{steel96});  
a nuclear emission region (B); two faint star-forming 
knots (C) associated with the tidal tail which extends to the south-west of 
the galaxy; and one or more faint knots to the south (D). 
These star-forming regions are labeled in the 4.5 \micron\ IRAC image of \hh\
(to be discussed later)    
in the left panel of Fig. \ref{fig:irac}.
Near-IR (NIR) and high-resolution VLA radio observations by \citet{johnson04}
demonstrate that the current star formation in \hh\ is most active in 
region A, 
with a star formation rate of $\sim$ 0.6 M$_\sun$ yr$^{-1}$ and a 
total mass of young stars of $\sim$ 10$^6$ M$_\sun$ within 
a region of $\le$ 0.1 kpc radius. Region A is also the 
youngest star-forming region in \hh, with clusters of ages $\la$ 5 Myr.
While the Balmer decrement yields an extinction \av\ = 0.5 mag, 
\citet{johnson04} found that the Br$\gamma$ and radio fluxes imply a larger extinction  
in region A, \av$\,\sim\,$ 8. 
As for the clusters in region B, they are slightly older, with 
ages $\sim$ 8--10 Myr \citep{johnson04}.
High-resolution \hst /WFPC2 F606W imaging of \hh\ by \citet{mgt}
shows that regions A and B are separated by a conspicuous dust
lane. 
The various star-forming regions 
are superimposed on an underlying low-surface-brightness component with 
somewhat irregular contours, so that \hh\ is of type iI\footnote{iI refers
to a complex central structure (star-formation zones not located at or near the 
center) and outer irregular isophotes.
The most common BCD morphology is iE in which the centers show
complex structure surrounded by outer circular or elliptical isophotes.} 
in the BCD morphological classification of \citet{loose85}.
Its redder color suggests that it is composed of older stars made in previous 
episodes of star formation. 
A tidal tail to the south-west is clearly seen in the \hst\ image of \citet{mgt},
suggesting that 
some of the present star formation may have been triggered by a merger with 
another dwarf galaxy or \hi\ cloud.  
\citet{johnson04} proposed that the \hi\ properties of \hh\
may be suggestive of a small-scale interaction, although higher resolution
\hi\ imaging would be necessary to confirm this.

Spectroscopy of region A by \citet{izotovthuan04} and 
\citet{thuanizotov05} gives oxygen abundances of 
12$+$log(O/H)\,=\,8.30$\pm$0.01 and 8.34$\pm$0.01 respectively, in good agreement with the value 
12$+$log(O/H)\,=\,8.37$\pm$0.06  obtained earlier 
by \citet{steel96}. If we adopt 12$+$log(O/H)\,=\,8.32 and a 
solar oxygen abundance of 12$+$log(O/H)\,=\,8.91 \citep{anders89}, \hh\ would have a
metallicity of about 1/4 that of the Sun.
If, instead, we adopt the new solar calibration of \citet{asplund05} with
12$+$log(O/H)\,=\,8.65, then \hh\ would have a metallicity of 1/2 solar.

We describe our new Spitzer observations and their reduction 
in Section \ref{sec:data}. 
In Section \ref{sec:spectra}, we present the spectroscopic results, and
discuss the PAH features and the IR fine-structure lines. 
We also compare the observed IR emission-line intensities 
with those predicted by a CLOUDY \citep{cloudy,ferland98} photoionization 
model based on the observed optical emission line fluxes. 
The IR morphology of the star formation in \hh\ is described 
in Section \ref{sec:morph}.
In Section \ref{sec:dust}, we construct 
a SED of \hh\ from 1\,\micron\ to 3.6\,cm 
based on our \spitzer\ and literature data.
Grain properties and characteristics of
the brightest emission region are derived by fitting the SED
with a \dusty\ model \citep{elitzur,dusty}.
We discuss \hh\ in a general context in Section \ref{sec:discussion}
and summarize our conclusions in Section \ref{sec:conclusions}.

\section{\label{sec:data}Observations and Data Reduction}

As part of our Cycle 1 \spitzer\ proposal (PID 3139: P.I. Thuan), 
we have observed
\hh\ with IRS \citep{houckirs} in the 
low- and high-resolution modules (SL, SH, LH); 
with IRAC \citep{fazioirac} at 4.5 and 8\,\micron; 
and with MIPS \citep{riekemips} at 24, 70, and 160\,\micron.
The data were acquired over a period of several months on 17 Nov 2004 (IRS),
23 Nov 2004 (IRAC), and 3 Apr 2005 (MIPS).
For all instruments, we 
started the data reduction with images processed by 
the \spitzer\ Science Center (SSC) pipeline, 
i.e. the Basic Calibrated Data ({\it bcd}) and the corresponding masks 
(the DCE masks).
The masks flag potential spurious features in the images, such as 
strong radiation hits, saturated pixels, or nonexistent/corrupted data.

\subsection{\label{sec:irsdata}IRS Spectra}

\begin{figure}
\vspace{0.2cm}
\hbox{ \includegraphics[angle=0,width=0.5\linewidth]{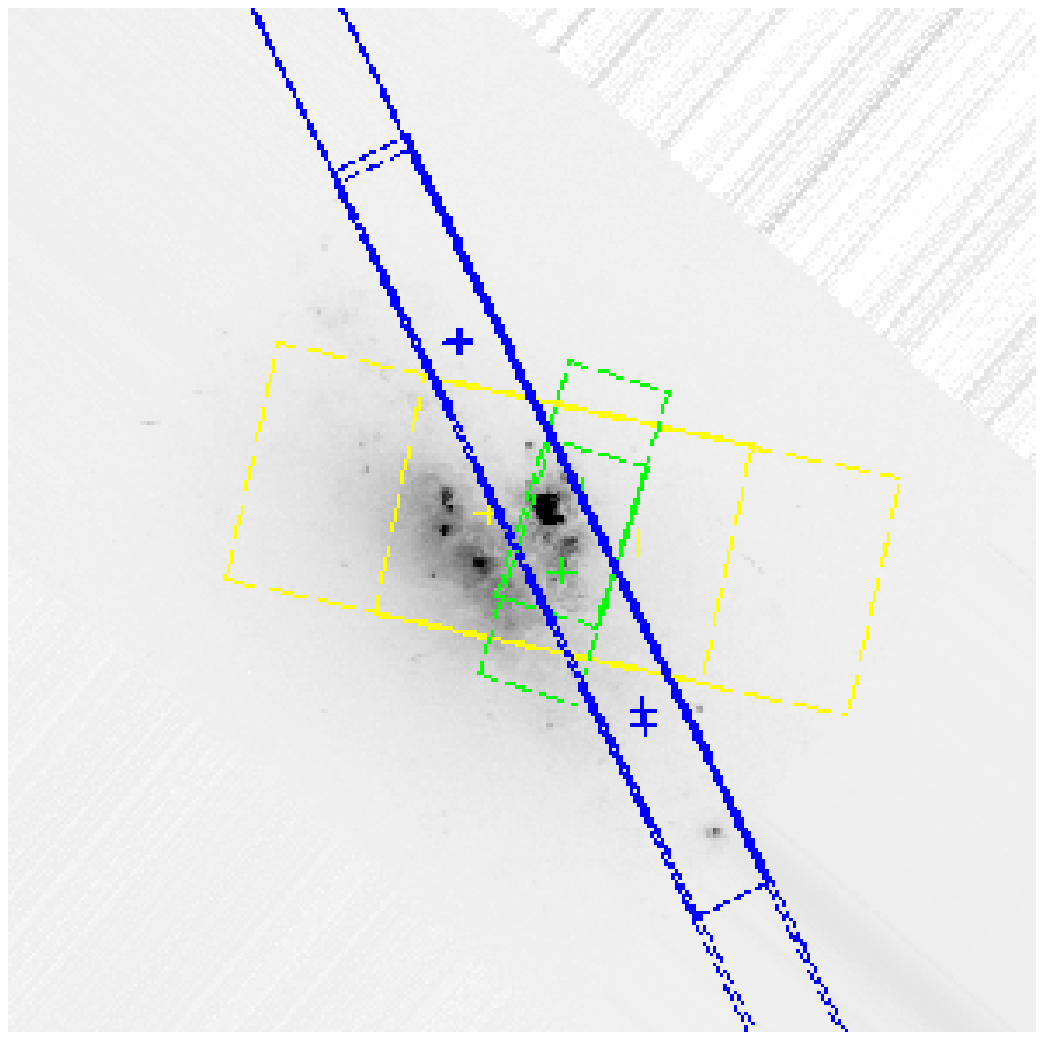}
\includegraphics[angle=0,width=0.47,width=0.5\linewidth]{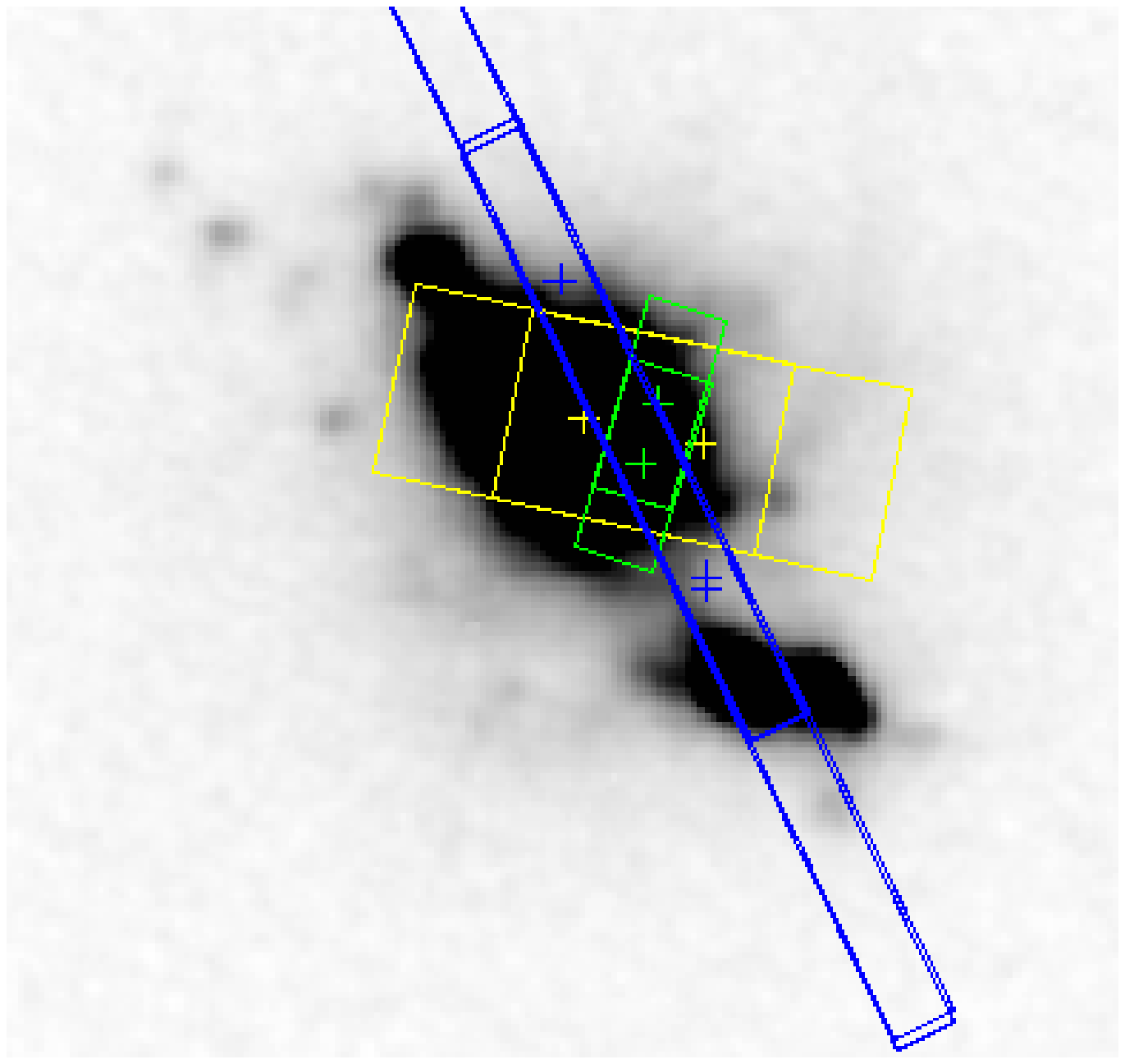} }
\caption{The IRS slit positions overlaid on the \hst /WFPC2 image (left panel,
\citet{mgt}) and the \ha\ image (right panel, \citealt{gildepaz03}) of \hh.
Images are oriented North up, East left;
the long (SL) IRS slit is $57\arcsec\times3\farcs6$ (blue), 
the SH slit $4\farcs7\times11\farcs3$ (green), 
and the LH slit $11\farcs1\times22\farcs3$ (yellow).
\label{fig:slits}}
\end{figure}

Spectroscopy was performed in staring mode with the Short Low module in both orders (SL1, SL2)
and with both Short and Long High-resolution modules (SH, LH) \citep{houckirs}.
These give 
low-resolution spectra from 5.2 to 14.5\,\micron\ (R$\simeq$64-128), and 
high-resolution spectra from 9.6 to 37.2\,\micron\ (R$\simeq$600).
We obtained 30s$\times$4 cycles for SH, 14s$\times$8 for LH,
6s$\times$3 for SL1, and 6s$\times$6 for SL2.
Sources were centered in the slits by peaking up and offsetting 
from 2MASS stars. 
Figure \ref{fig:slits} shows the positions of the three 
different slits across \hh;
in the left panel, the slits are superimposed on the \hst/WFPC2 F606W image \citep{mgt}
and in the right panel on the \ha\ image of \citet{gildepaz03}. 
It can be seen that all three slits include region A. 

It is necessary to accurately 
subtract the background before spectrum extraction. 
Because of the dominance of the background signal,  
the accuracy of this process strongly influences the quality of the 
final spectrum.
We started the spectral reduction with
the individual {\it bcd} frames processed by the S11.0.2 version of the 
SSC pipeline, which provides ramp fitting, dark current 
subtraction, droop and linearity corrections,
flat-fielding, and wavelength and flux calibrations\footnote{See the IRS
Data Handbook, \url{http://ssc.spitzer.caltech.edu/irs/dh}.}.
However, the pipeline does not include background subtraction.
Hence, for the low-resolution spectra, we constructed a coadded 
background frame from the 
{\it bcd} observations 
with the source in the opposing nod and off-order positions
\citep[see also][]{weedman05}. 
When the source is in SL1, the SL2 slit samples background, and vice versa.
This makes it possible to include the off-order frames in the background image
since the same on-chip integration times were used for both SL orders.
The image for the source at a given nod position was constructed by 
coadding all frames at that position.
Coadding was performed with the sigma-clipping option
of the {\it imcombine} task
in IRAF\footnote{IRAF is the Image Analysis and 
Reduction Facility made available to the astronomical community by the National Optical
Astronomy Observatory, which is operated by AURA, Inc., under
contract with the U.S. National Science Foundation.}. 
The inclusion of off-order and off-nod frames in the background image 
means that
the integration time on the background is three times that on the source, which
improves the signal-to-noise of the two-dimensional (2D) subtraction.

For the high-resolution SH and LH spectra, 
a 2D background image could not be 
constructed because of the small size of the slit.
Therefore we subtracted the background from the SH and LH observations
using the one-dimensional (1D) spectra as described below.
The SH and LH 2D {\it bcd} images were coadded as for the SL modules,
and successive corrections for sporadic bad pixels and cosmic ray
hits were carried out manually by inspection of the images at
the separate nod positions.

We extracted the source spectra with \spice, the post-pipeline IRS package
furnished by the SSC.
To maximize the calibration accuracy,
the automatic point-source extraction window was used for all modules.
The automatic extraction uses a variable-width extraction window which
scales with wavelength in order to recover a constant fraction of 
diffraction-limited instrument response.
For the SL spectra, this gives a 4-pixel (7\farcs2) length at 6\,\micron, and
an 8-pixel one (14\farcs4) at 12\,\micron; 
the slit width is 3\farcs6 for both SL modules.
At high resolution, the \spice\ extraction is performed over the entire slit
(4\farcs7$\times$11\farcs3 SH; 11\farcs1$\times$22\farcs3 LH).
Orders were spliced together by averaging, ignoring the noisy
regions at the red end of each order \citep[e.g.][]{armus04}.
Then the individual spectra were box-car smoothed to a resolution element, 
and clipped in order to
eliminate any remaining spikes in the high-resolution data.
Finally, the two spectra for each module (one for each nod position) 
were averaged.

Background was subtracted from the 1D SH spectra by minimizing the
difference between the 2D background-subtracted SL and SH spectra over 
the substantial ($\sim$5\,\micron) overlap in wavelength. 
We adopted the model of Reach and coworkers\footnote{See
\url{http://ssc.spitzer.caltech.edu/documents/background}.} for
the spectral shape of the background,
and the multiplicative constant was given by the minimization. 
The LH background was subtracted by minimizing the difference
between the SH and LH spectra over their overlap region ($\la$1\,\micron).

The  final averaged IRS 
spectra obtained in 
the short-wavelength low-resolution mode (SL) and in both short and 
long wavelength  
high-resolution modes (SH, LH) are shown in Fig. \ref{fig:irs}. 

\subsection{\label{sec:iracdata}IRAC Images}

We designed our 
IRAC Astronomical Observation Requests (AORs) to give  four (high-dynamic 
range\footnote{This AOR option gives two sets of four frames each: one with
the specified integration time, and one with a shorter time, typically $\sim$1s.}) 
IRAC frames of 30s; only channels 2 (4.5\,\micron) and 4 (8.0\,\micron) were
acquired. 
Small-scale dithering in a ``cycling'' pattern was performed with
a total of 120s spent in each IRAC channel.

The individual {\it bcd} frames were processed with  
the S11.0.1 version of the SSC pipeline
(see the IRAC Data Handbook\footnote{Available from the SSC website
\url{http://ssc.spitzer.caltech.edu/irac/dh/}}).
This processing is designed to remove the effects of
dark current, detector nonlinearity, flat field, multiplexer bleeding, 
and cosmic rays, and to perform flux calibration.
However, in some cases, those with particularly bright sources in 
the field-of-view, 
the pipeline is unsuccessful at removing all artefacts.
In the case of \hh, the bright point-like source at the center of the
galaxy resulted in a ``banding'' effect in the 8\,\micron\ image, i.e.  
in an artificially enhanced brightness level in the rows and columns 
containing that bright source (see the IRAC Data Handbook).
The 4.5\,\micron\ image was not affected by banding since it occurs only in the
Si:As arrays (channels 3 and 4).
Dithering does not eliminate the effect because the position of the
bands on the array remains unchanged relative to the galaxy.
Moreover, because the intensity of the brightness enhancement decreases with 
distance from
the bright source, it is not straightforward to correct automatically.
We eliminated the effect from each 8\,\micron\ frame by interpolating 
the feature across the affected rows, 
substituting the spurious pixel values with those of a typical background.

The {\it bcd} frames (corrected for ``banding'' effects in the case of the 
8\,\micron\ ones)
were coadded using \mopex, the image mosaicing and source-extraction package
provided by the SSC \citep{mopex}.
Pixels flagged by the DCE masks and by the static bad-pixel masks (pmasks) were
ignored.
Additional inconsistent pixel values were removed by means of the \mopex\ outlier
rejection algorithms. 
We relied on the dual-outlier technique, together with the multiframe algorithm,
rather than using any spatial criteria for rejection.
The frames were corrected for geometrical distortion and
projected onto a ``fiducial'' (refined) coordinate system
with pixel sizes of 1\farcs20, roughly equivalent to the original pixels.
Standard linear interpolation was used for the mosaics.
The noise levels in our post-pipeline \mopex\ mosaics are comparable to
or lower than those in the SSC products.
The final coadded 4.5 and 8.0\,\micron\ images are shown 
respectively in the left and right panels of Fig. \ref{fig:irac}.
We also show 4.5\,\micron\ contours superimposed on 
the \hst/WFPC2 F606W image of  
\citet{mgt} in Fig. \ref{fig:iracoverlay}.
Examination of Fig. \ref{fig:iracoverlay} shows that 
the 4.5\micron\ peak coincides with the bright optical point source
associated with the compact radio emission \citep{johnson04}.

\subsection{\label{sec:mipsdata}MIPS Images}

Our MIPS images were acquired in the Fixed Cluster-Offset mode in all
three channels, with offsets of 12\arcsec\ in two additional
pointings.
We used ramp times of 3s, 10s, and 10s for 
24\micron, 70\micron, and 160\micron, respectively, with 1, 2, and 6 cycles in the
three channels.
This gave us a total of 32 frames at 24\micron, 56 frames at 70\micron, 
and 104 frames at 160\micron.
Because the first data frame of each observation sequence at 24\micron\ has a shorter 
exposure time, we discarded those frames with keyword DCENUM\,=\,0;
we also removed the second frame (DCENUM=1) because of 
increased data artefacts 
(see the MIPS Data Handbook\footnote{Available from the SSC website
\url{http://ssc.spitzer.caltech.edu/mips/dh/}}); 
this left us with 24 frames at 24\micron.

\begin{figure}
\vspace*{-15mm}
\hbox{\hspace{-0.5cm}\includegraphics[angle=270,width=0.5\linewidth]{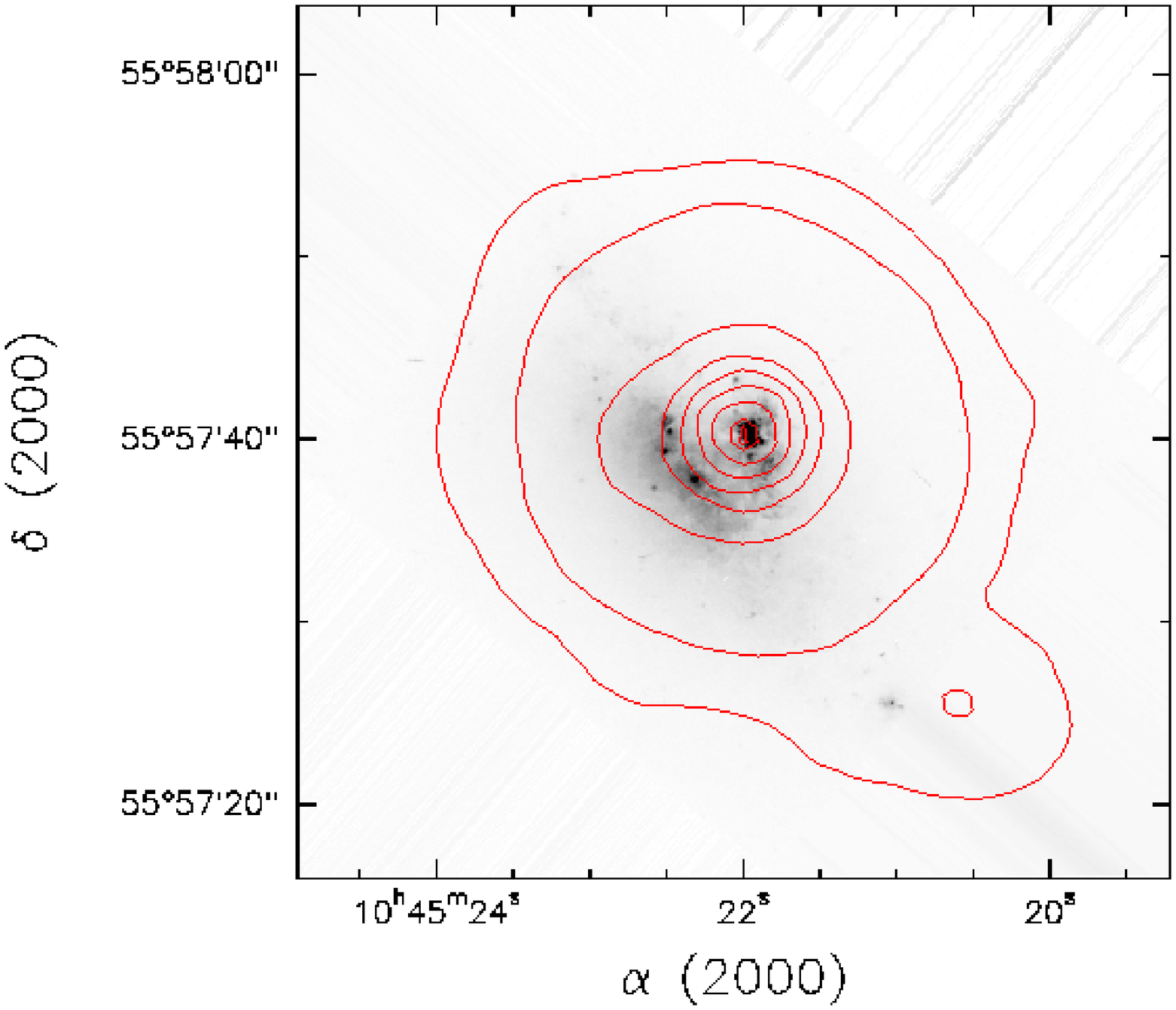}
\includegraphics[angle=270,width=0.45\linewidth]{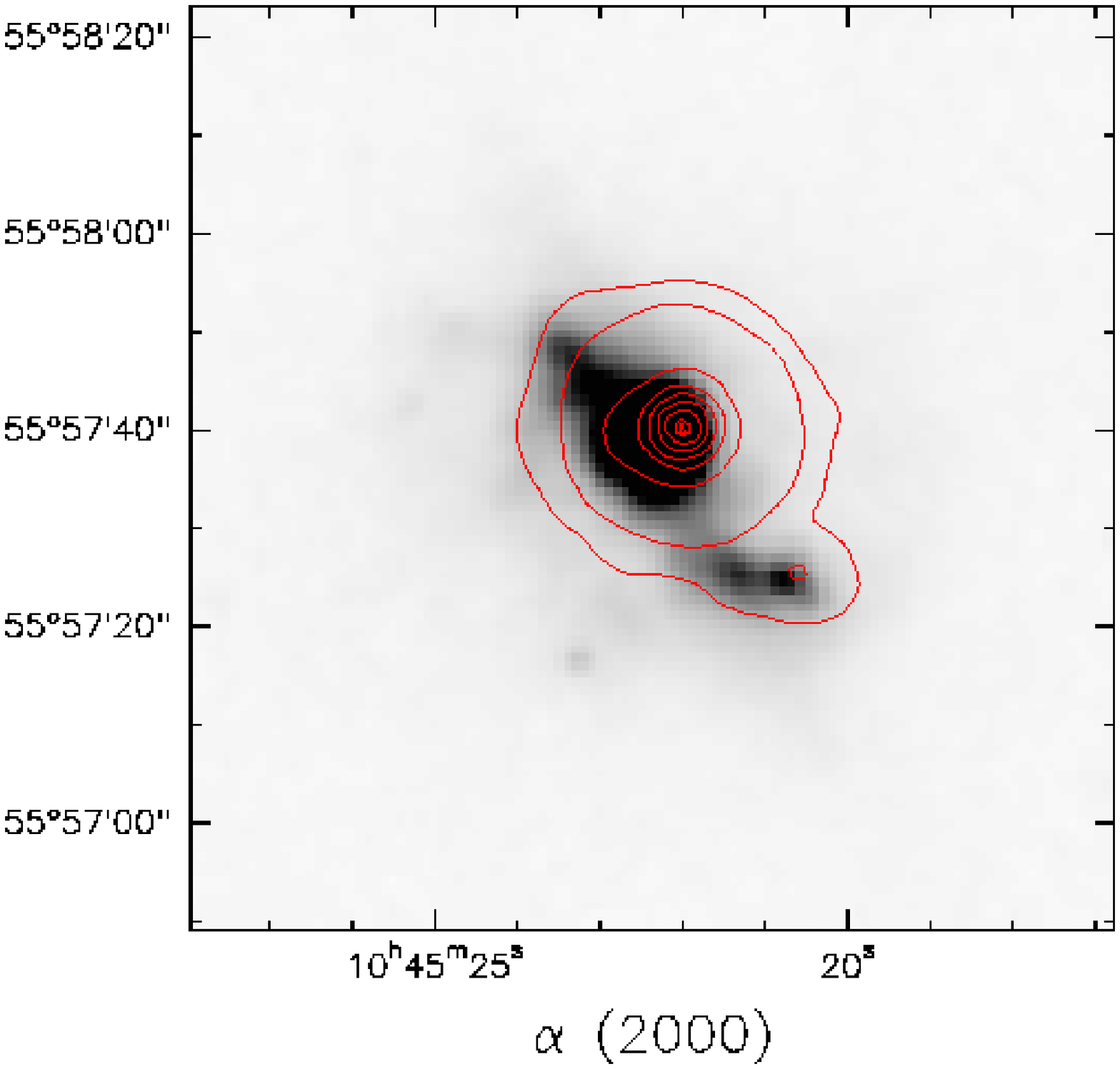} }
\hbox{\hspace{-0.5cm}\includegraphics[angle=270,width=0.5\linewidth]{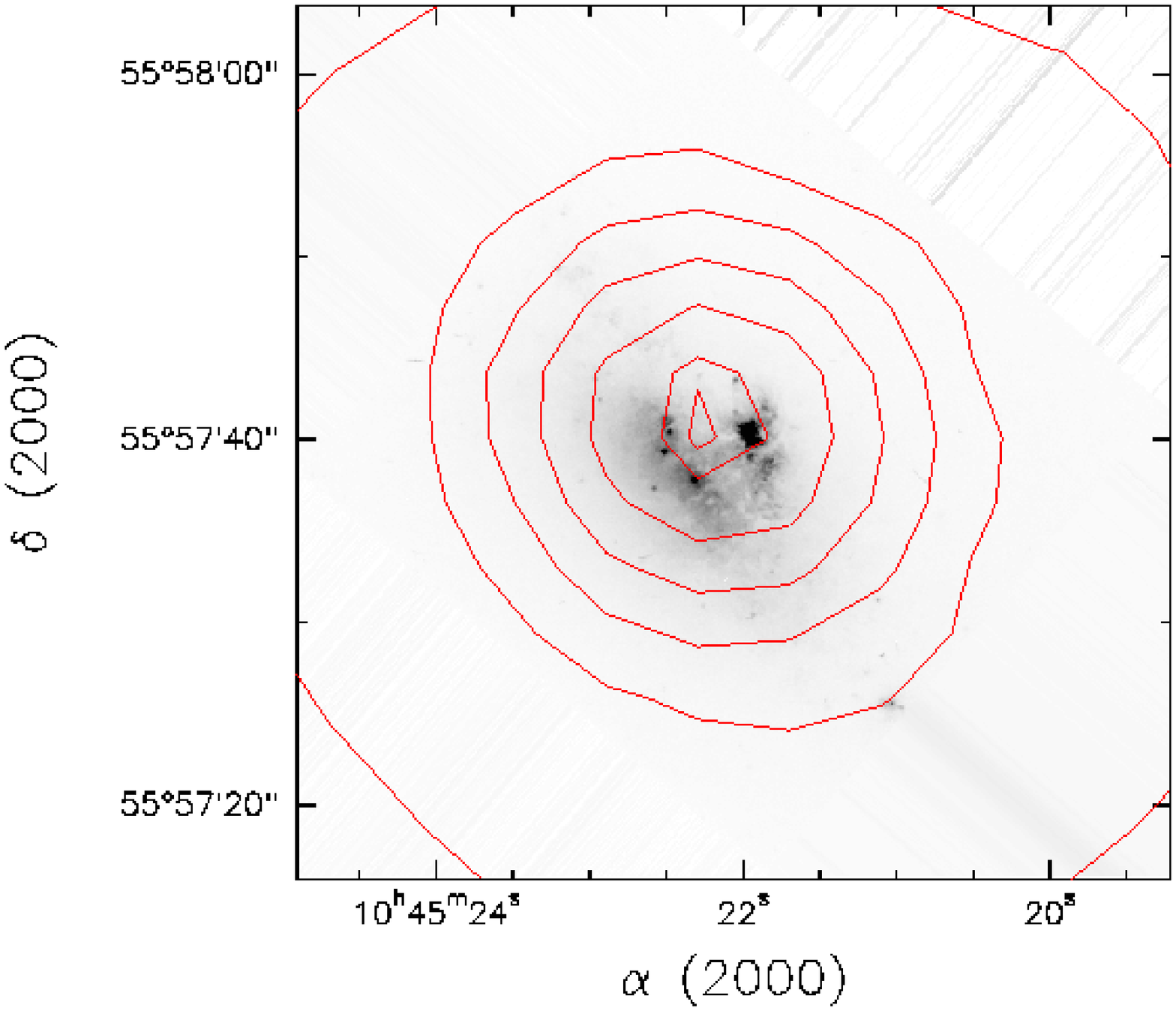}
\includegraphics[angle=270,width=0.45\linewidth]{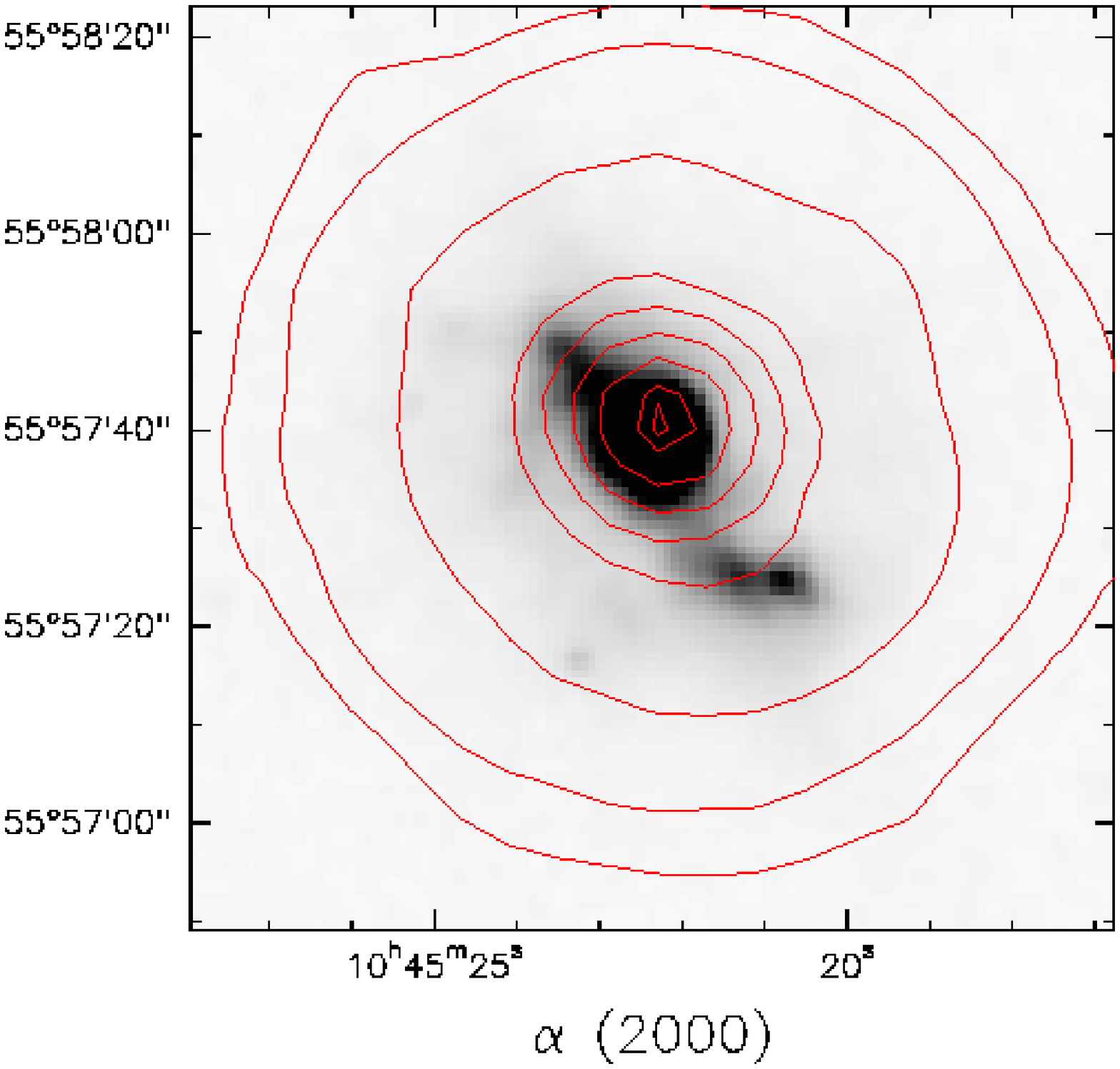} }
\hbox{\hspace{-0.5cm}\includegraphics[angle=270,width=0.5\linewidth]{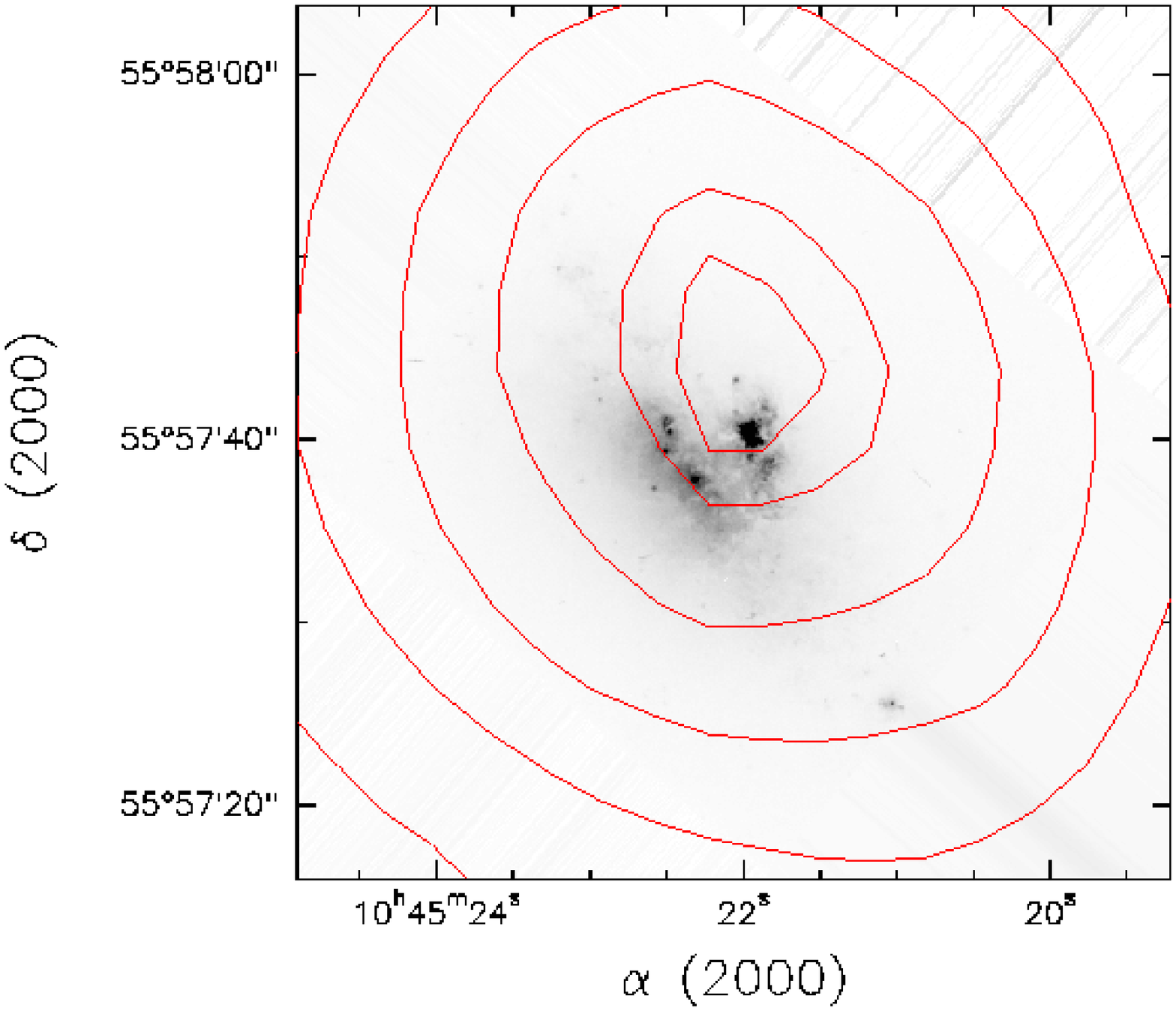}
\includegraphics[angle=270,width=0.45\linewidth]{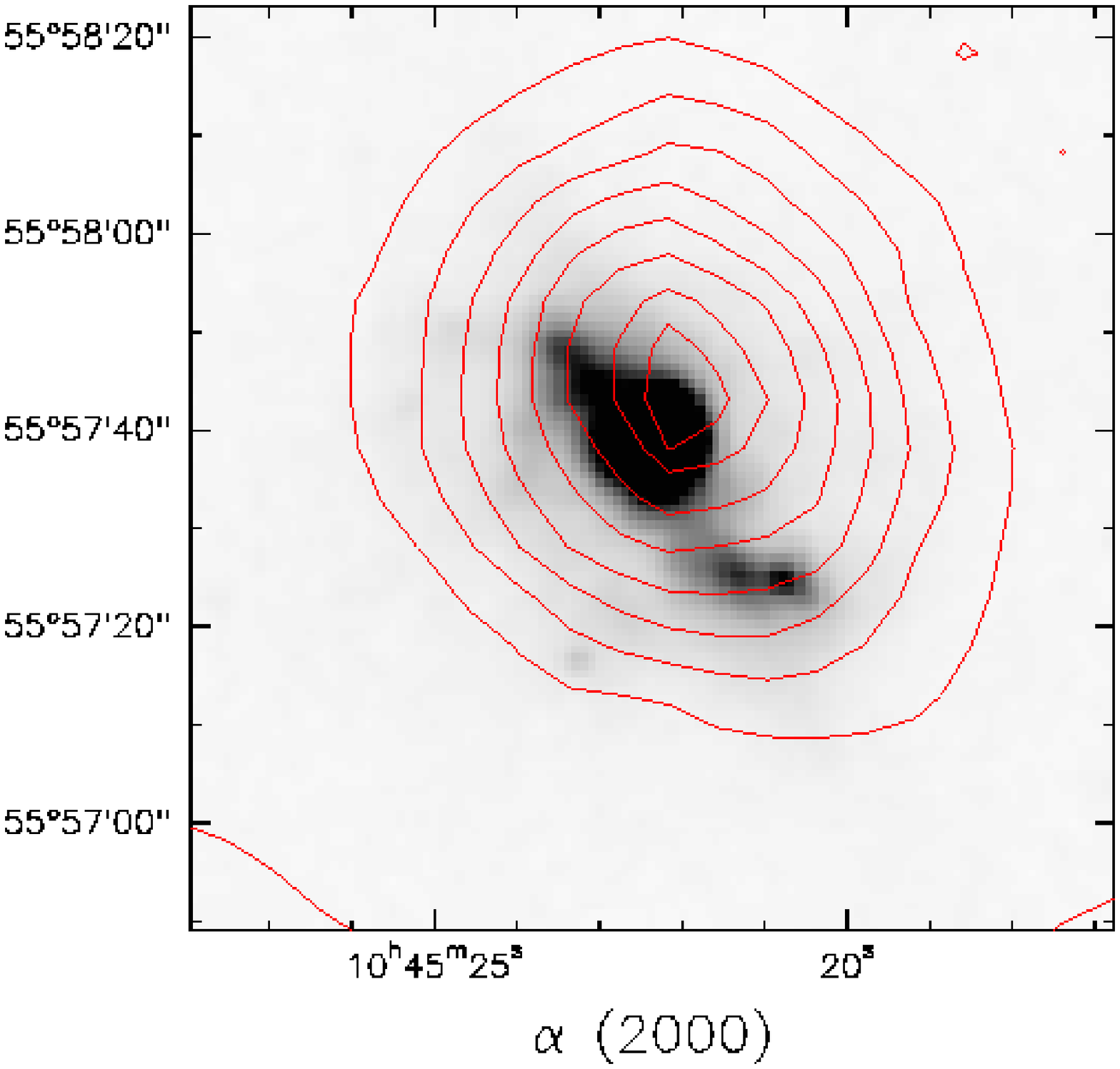} }
\caption{The 24 \micron\ (top panel), 70 \micron\ (middle), and 160 \micron\
(bottom) MIPS images of \hh\ 
contoured on the (left panels) \hst /WFPC2 F606W image \citep{mgt}, 
and on the 8.0\micron\ IRAC image (right).
Contours run 1\% above the sky, then 8\%, 15, 30, 45, 60, 75, 90, and 95\% 
of peak value, except for MIPS\,160, where they run from 10\% above the
sky, then 30 to 90\% (in increments of 10\%), and 95\% of peak value.
\label{fig:mipsoverlay}}
\end{figure}

The individual {\it bcd} frames were processed by the S11.4.0 version of the SSC pipeline,
which converts the integration ramps inherent to the MIPS detectors into 
slopes, and remediates temporal variation of the slope images
\citep{gordon05}.
The processing includes correcting for stimulator flashes on the MIPS-Ge detectors,
dark subtraction, flat-fielding, and flux calibration.
As for the IRAC images, we processed the dithered {\it bcd} frames in 
the spatial domain with \mopex.
The DCE masks and the static masks were used to flag pixels, and subsequently ignore them.
The \mopex\ outlier rejection was used to flag any additional spurious pixel
values, with the dual-outlier and multiframe algorithm as for the IRAC frames. 
Geometrical distortion was corrected before projecting the frames onto 
a fiducial coordinate system with pixel sizes of 1\farcs20 for MIPS-24,
roughly half of the original pixel size of 2\farcs5.
Pixel sizes of the final mosaics at 70 and 160\micron\ are also 
approximately half
of the originals, i.e. 4\farcs95 at 70\micron\ and 8\farcs0 at 160\micron.
Unlike the IRAC coadds,
we incorporated the sigma-weighting algorithm because we found
it gave less noisy MIPS mosaics than without.
Standard linear interpolation was used in all cases.
In all three channels, our  post-pipeline \mopex\ mosaics are superior to 
those provided by the automated post-pipeline reduction.
The final coadded images are shown in Fig. \ref{fig:mipsoverlay} as 
contours superimposed on
the \hst/WFPC2 (left panel) and on the IRAC 8\,\micron\ images (right panel).

\subsection{\label{sec:phot} IRAC and MIPS Photometry}

\begin{figure}
\vspace{0.2cm}
\hbox{\hspace{-0.5cm}\includegraphics[angle=0,width=0.5\linewidth]{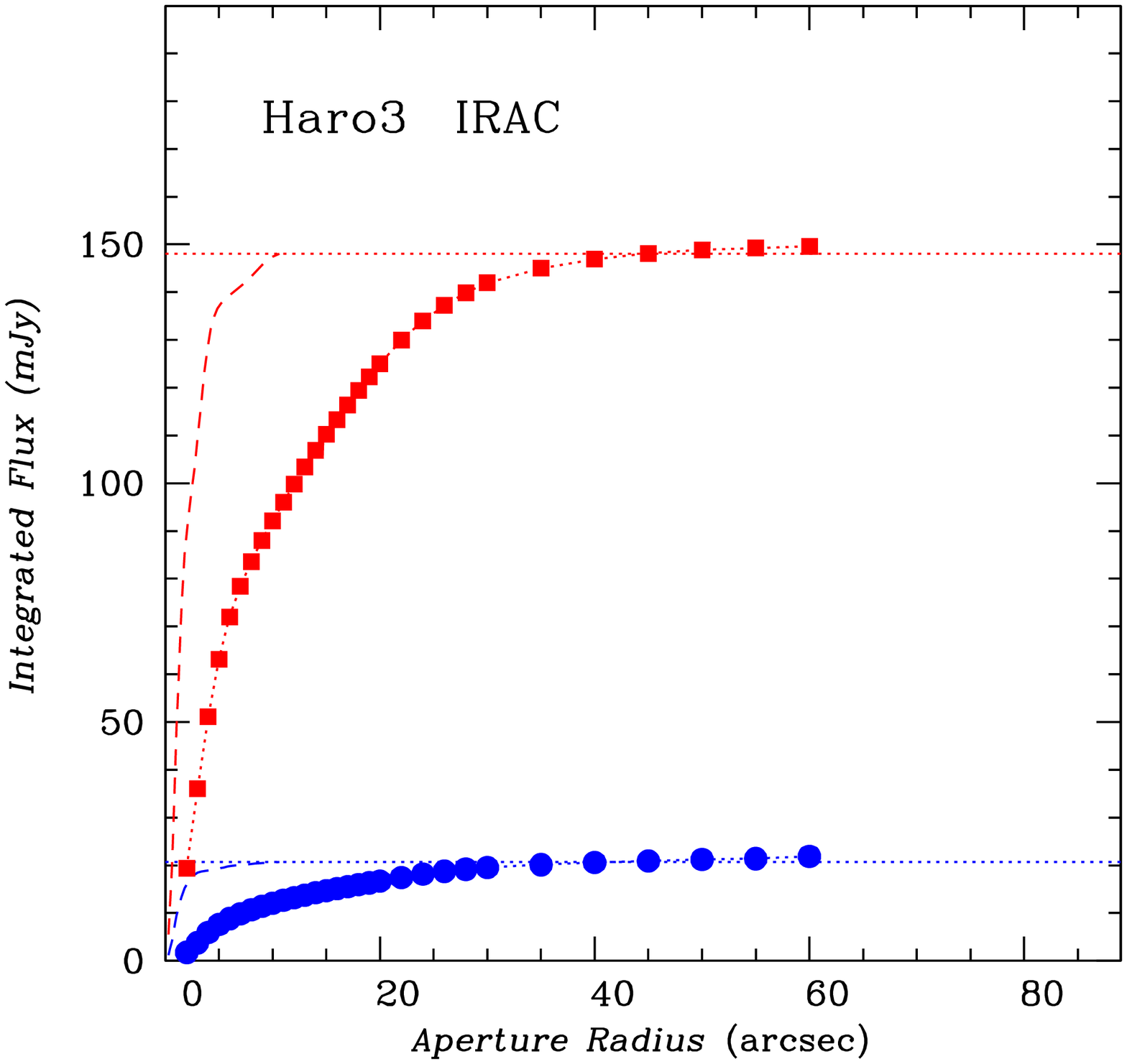}
\includegraphics[angle=0,width=0.5\linewidth]{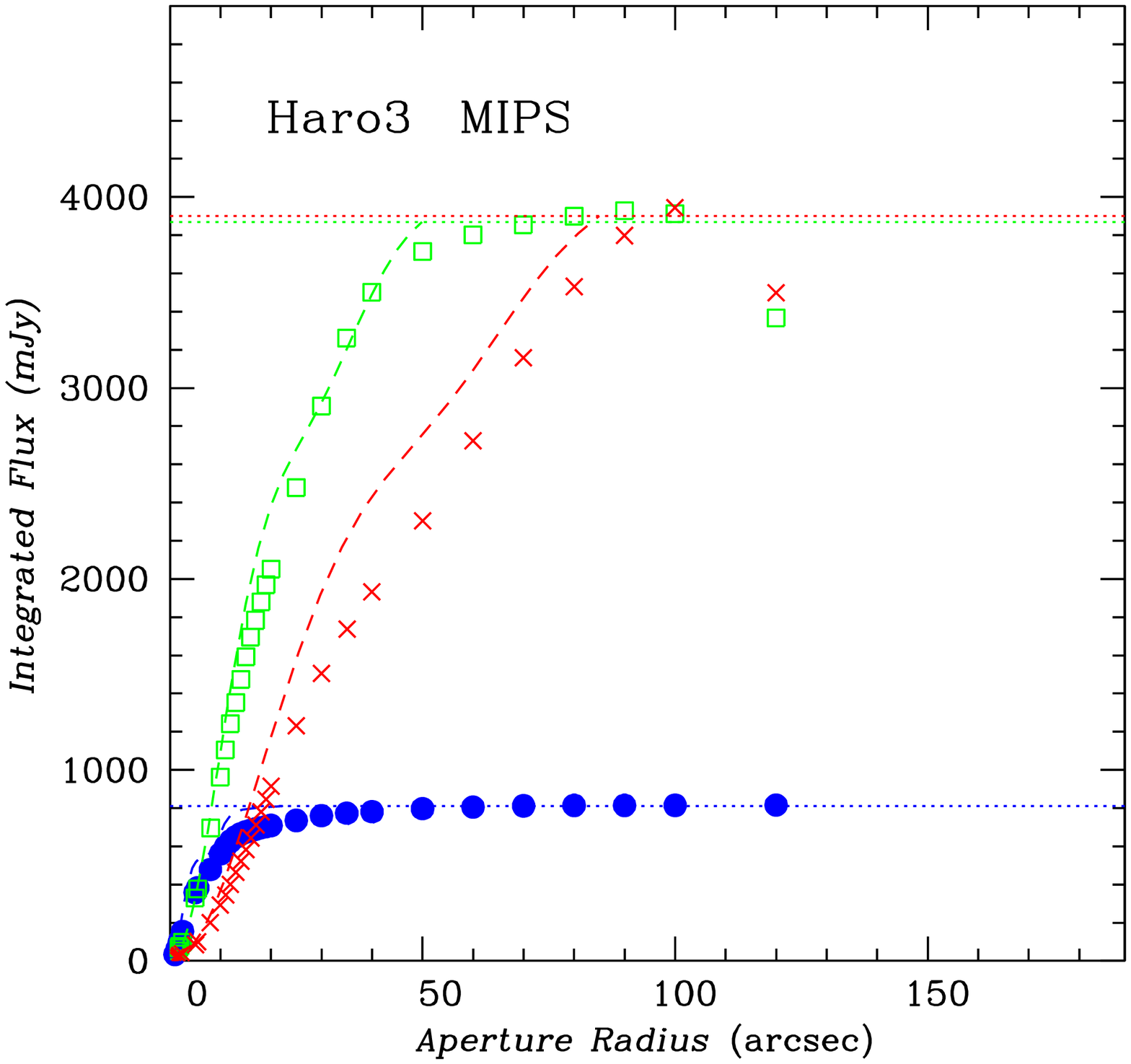} }
\caption{Growth curves for IRAC (left panel) and MIPS (right) photometry. 
In both panels, the horizontal dotted line traces the adopted total
flux from \hh, and
the dashed line represents the point-response function.
In the left panel, IRAC channel 2 is indicated by filled circles,
channel 4 by filled squares.
In the right panel, MIPS-24 is shown by filled circles,
MIPS-70 by open squares, and MIPS-160 by $\times$.
\hh\ is extended at shorter wavelengths and possibly at 160\micron, but
appears virtually point-like at 24 and 70\micron. 
\label{fig:phot}}
\end{figure}

We have performed aperture photometry on the IRAC and MIPS images
with the IRAF photometry package {\it apphot}, taking care to convert
the MJy/sr flux units of the images to integrated flux.
The background level was determined by averaging several 
adjacent empty sky regions. 
The photometric growth curves are shown in Fig. \ref{fig:phot}, for  
IRAC in the left panel, and for MIPS in the right panel.
We have also made photometric measurements 
 of the instrument point response functions (PRFs),
provided in the \mopex\ package\footnote{We used the most recent determinations
of the PRFs dated 5 October 2005.}; 
the dashed lines show the growth curve expected from a point source, normalized
to the total flux indicated by horizontal dotted lines in Fig. \ref{fig:phot}.

Inspection of Fig. \ref{fig:phot} shows that \hh\ is clearly extended at IRAC
wavelengths (4.5 and 8\,\micron), and possibly at 160\micron, but 
appears virtually point-like at 24 and 70\,\micron.
The morphology of the 24\,\micron\ image is essentially that of a point source
(see Fig. \ref{fig:mipsoverlay}).
To check for source extension, we have fitted elliptical 
gaussians to the brightest source in the IRAC and MIPS-24 images.
The width (equal to the gaussian $\sigma$) of these fits is $\ga$\,2\farcs4 
in both the IRAC 
and MIPS-24 images,
which implies that, indeed, the source is quite extended at 
IRAC wavelengths, but
more point-like at 24\micron.
However, at 24\,\micron, the diffraction limit is $\sim$6\arcsec, 
implying that 
the gaussian fit is too narrow;
moreover the 24\micron\ flux resulting from this fit may be too low 
(see $\S$\ref{sec:dust}), because the fit does not account
for the substantial flux in the first Airy ring. 
The fraction of total flux contained in the gaussian-like source is 40\%, 30\%,
and 58\%, at 4.5, 8.0, and 24\,\micron, respectively.
The total IRAC and MIPS fluxes are given in Table \ref{tab:photom}, along 
with other published near-infrared, far-infrared and radio photometry on \hh.
For the \spitzer\ photometry, we adopt a global uncertainty of 10\%, which should
account for the various contributions to the uncertainties, including flux
calibration and imperfect sky subtraction.

\subsection{\label{sec:compare}Comparison With Previous Work}

\hh\ was observed with the {\it Infrared Space Observatory} (\iso) by \citet{metcalfe96} 
and by the {\it Infrared Astronomy Satellite} (\iras) (Table \ref{tab:photom}).
The flux measured in the \iso\ LW6 filter (7-8.5\micron) of 149$\pm$30\,mJy
is in excellent agreement with the total flux of 148\,mJy measured 
in the 8\micron\ channel of IRAC.
The total flux of
300$\pm$70\,mJy given by \citet{metcalfe96} in the \iso\ 12-18\micron\ 
LW3 filter is also consistent with the IRS spectrum, considering
that \hh\ is slightly extended at these wavelengths.
The IRS spectrum levels generally agree well
with other measurements, implying that the background subtraction is reliable.

Our measured \spitzer\ 24\micron\ total flux of 810$\pm$81\,mJy agrees well with the
\iras\ flux of 942$\pm$57\,mJy at 25\micron.
The MIPS long-wavelength fluxes at 70 and 160\micron\ of $\sim\,$3.9 Jy 
are slightly lower than the \iras\ fluxes of 4.9 (60\,\micron) and 6.7 (100\,\micron) Jy
taken from NED\footnote{The NASA/IPAC Extragalactic Database (NED) is operated 
by the Jet Propulsion Laboratory, California Institute of Technology, under 
contract with the National Aeronautics and Space Administration.}. 
But given that they are measured in smaller apertures, our fluxes are 
consistent with those of IRAS.  We conclude that our measurements are in 
general good agreement with previous existing data and that we can have 
confidence in our reduction procedures. We next discuss how our
Spitzer 
data, combined with optical, near-infrared and radio data, can 
constrain the properties of the stellar populations, the ionized gas and 
the dust, and help us to construct a picture of the star formation in \hh.

\section{\label{sec:spectra}Spectroscopic results}

Polycyclic Aromatic Hydrocarbon (PAH) emission from small dust grains and
gaseous nebular line emission are visible in the IRS spectrum of \hh\
(Fig. \ref{fig:irs}).
These features are superimposed on a steeply rising continuum which peaks
around $\sim$100\,\micron\footnote{This is where the flux
distribution peaks (see Table \ref{tab:photom}); the SED (energy
distribution) peaks at $\sim$40\,\micron.}. 

\subsection{PAH Features}
Examination of the IRS spectra in  Fig. \ref{fig:irs} 
shows that there are PAH
features clearly detected at 5.7, 6.2, 7.7, 8.6, 11.2, 
12.7, and 16.4\,\micron. We have derived 
the flux and equivalent width (EW) of these features by
fitting the SL and SH spectra with a series of lorentzian profiles of unknown 
width, amplitude, and central wavelength. 
Because of their broad low-intensity wings, the PAH
features are better fit with lorentzian than with
gaussian profiles.
The fits were performed with the {\it splot} task in IRAF.
The continuum was linearly interpolated in two sections, one
from 6-9\,\micron, and the other from 10.5-13.9\,\micron. 
The best-fit parameters, averaged over several independent estimates
in both high and low resolution spectra, are given in Table \ref{tab:pahs}.
The flux of the 12.7\,\micron\ PAH feature is probably overestimated
because of the difficulty in separating it from the bright \neii\ 
fine structure line (see $\S$\ref{sec:lines}).

Our derived values are in general agreement with those based on 
ISO spectra \citep{metcalfe96}. 
There is however a large difference in the slope of the continuum 
between the ISOPHOT and IRS spectra. 
The ISOPHOT continuum is nearly flat while 
the IRS one is continuously rising toward longer wavelengths 
(Fig. \ref{fig:irs}).
We interpret this continuum slope difference as due to the fact 
that different regions in \hh\ are sampled by ISOPHOT and IRS, 
as a result of the    
different apertures in the two instruments.  
The ISO PHT-S beam is $25\arcsec\times25$\arcsec ($\sim$ 2\,kpc),  
considerably larger than the IRS apertures. For the latter, 
the slit width is 3\farcs6 for both SL modules, while it is 
4\farcs7$\times$11\farcs3 for the SH module and 
 11\farcs1$\times$22\farcs3 for the LH module. This means that, while the 
IRS apertures include mainly radiation from the 
star-forming region A, the ISO aperture encompasses a large 
part of the interstellar medium (ISM) around region A. 
So the spectrally flat PAH emission seen by
ISOPHOT comes from the general extended ISM in \hh\ while 
the strong rising MIR continuum emission seen by IRS is related to the 
more compact star-forming region A.          

The relative strengths of the three main PAH emission features can
be used as diagnostics to
identify PAH sizes and infer the neutral-to-ionized gas ratio
\citep{draine01}.
According to the models of \citet{draine01} and the
observed flux ratios of the $f_{11.2}/f_{7.7}$ and $f_{6.2}/f_{7.7}$ features, 
the PAHs in \hh\ are predominantly neutral and small, containing a 
few hundred carbon atoms, similar to the 
normal galaxies studied by \citet{helou00}. In other words, \hh\ lies in 
the ``normal galaxy'' region in the PAH diagnostic diagram rather than in the
starburst galaxy region which includes objects such as M\,82 and NGC\,253,
that are dominated by ionized PAHs.
This again supports our previous contention that the PAHs in \hh\ arise 
mainly from the normal ISM around the star-forming region, 
not only from knot A itself.

On the other hand,
the PAH EWs are generally high, $\sim 2.9$ for the 7.7\,\micron\ line, 
more typical of starburst galaxies \citep{brandl04} than of
BCDs \citep{wu06}. This result may not be surprising as \hh\ with M$_B$\,=\,$-17.9$ 
is at the bright end of the luminosity function of dwarf galaxies,
but it contrasts with the flux ratio trends described above.
Overall, \hh\ follows the metallicity trends of
increasing PAH EW with increasing oxygen abundance reported by \citet{wu06} 
and discussed by \citet{hogg05} and \citet{engelbracht05}.
However, for its metallicity, 
the EWs of the ionized PAHs in \hh\ exceed by $\sim$50\% to up
to almost a factor of three that of 
most extreme BCD in the \citet{wu06} sample.
In contrast, the neutral PAH at 11.2\,\micron\
has an EW of 0.6, typical of other BCDs.

Finally, we find that the PAH emission features in \hh\ are narrow, 
narrower than many of the objects tabulated by \citet{li01}.
The only objects which possess PAH features with 
smaller widths at 7.7\,\micron\ are the Small Magellanic Cloud, 
the reflection nebulae/photodissociation region NGC\,7023,
and 30\,Doradus in the Large Magellanic Cloud.
M\,82 and NGC\,253 have slightly broader PAH features than
these objects or \hh, but the difference may not be significant. 
However, different PAH features tend to have systematically
different FWHMs \citep{li01}, so that to make a more definitive
statement, we must wait for more data and a more uniform method
of measuring PAH parameters \citep[e.g., {\it PAHFIT}\,:][]{pahfit06}. 
Perhaps not surprisingly,
\hh 's PAH features are also narrower than those predicted by 
the model for the
diffuse Galactic ISM at high latitudes \citep{li01}.
We have tried to subtract a scaled version of the model spectrum from 
the observed spectrum of \hh, but because of the narrowness of the 
features and their lower equivalent widths (because of the rising
continuum), the scaled subtraction leaves significant residuals.
We conclude that there are many factors behind the appearance of 
PAH features (e.g., chemical abundance, ISM energetics, etc.).
Thus even though our IRS spectrum may contain a conspicuous
ISM component, Galactic ISM spectra may not be a good template
for all ISM PAH emission.

\subsection{Infrared Fine-Structure Lines \label{sec:lines}}

The spectrum also shows several fine-structure lines, including
the highest signal-to-noise detections \siv $\lambda$10.51, \neii $\lambda$12.81,       
\neiii $\lambda$15.55, 36.01, \siii $\lambda$18.71, 33.58, \oiv $\lambda$25.89, and 
\SiII $\lambda$34.82\,\micron.
As for the PAH features, we used
{\it splot} to fit the IR fine-structure emission lines and obtain
fluxes and EWs.
A deblending procedure was adopted to accurately measure
the emission lines at wavelengths near PAH features
(e.g., \ariii\ and \neii).
Unlike the PAHs, the lines were best fit with a gaussian profile.
A single continuum was linearly interpolated from 10.5 to $\sim15$\,\micron.
For the longer wavelengths, the local continuum was fit by linearly
interpolating over adjacent line-free regions. 
The fluxes and other parameters resulting from the fits
are reported in Table \ref{tab:lines}.

Several fairly high-excitation lines are detected in \hh.
The \oiv\ line at 25.89\,\micron\ is faint, but present.
\oiv\ is a high excitation line, with an excitation potential of 
54.9\,eV, just beyond the \heii\ edge at 54\,eV.
Its detection is consistent with the Wolf-Rayet population present
in \hh 's starburst \citep{steel96,thuanizotov05}.
\citet{thuanizotov05} have looked in the optical spectrum of \hh\
for the high-ionization \fev\ $\lambda$4227 
and \heii\ $\lambda$4686 nebular emission lines which both have an 
ionization potential of 54.4\,eV.
None were seen. 
Evidently, the observed optical ionizing radiation is less hard than that in 
the MIR range. This may suggest the presence of substantial extinction in the blue,
or indicate that the optical and MIR radiations come
from different locations in the star-forming region A, or both    
(see also $\S$\ref{sec:cloudy} and $\S$\ref{sec:dust}).

Line ratios of different ionic species of the same element,
e.g., \neiii\ and \neii, are sensitive to the shape of
the ionizing radiation field. 
The ratio of the \neiii\ and \neii\ lines is 2.8, placing \hh\
toward the high-excitation end of the starbursts studied by
\citet{verma03}, and similar to NGC\,5253. 
Because PAHs tend to be depleted in hard radiation fields,
their EW would be expected to be negatively correlated with 
the \neiii/\neii\ line ratio. 
Such a trend was indeed found by \citet{wu06} for their BCD sample. 
However, compared to the \citet{wu06} correlation,  
the PAHs in \hh\ tend to have higher EWs for their metallicity than other BCDs.
This may again be due to a significant
PAH component from the general ISM in \hh\ as compared to the BCDs studied 
by \citet{wu06}, as 
the region of \hh\ encompassed by the IRS slit is relatively large, 
about 500\,pc.
It could also arise from different measurement techniques, as \citet{wu06}
fit a spline to the underlying continuum, while we linearly interpolate over
two sections of ``global'' continua.

We can estimate the electron density with the temperature-insensitive
\siii\ lines. 
The flux ratio $f_{33.84}/f_{18.71}$ is 2.5, well into 
the low-density limit \citep{draine06}.
The implied electron density is $\simlt$100\,cm$^{-3}$, lower than 
the optically-derived value of 180$\pm$30\,cm$^{-3}$ by \citet{izotovthuan04}
from the \sii\ $\lambda$6717/$\lambda$6731 ratio. 
Even with a maximum uncertainty on the \siii\ flux ratio of $\sim$20\%, these
two densities are inconsistent. 
Beam dilution could be affecting the MIR line ratios,
since the IRS aperture is several times larger than the optical one.
Alternatively, this discrepancy could be again suggesting
that the optical emission comes from a different spatial location 
in the star-forming region A as compared to the MIR emission.



\subsection{CLOUDY Models \label{sec:cloudy}}

Since the amount of hard radiation and the electron number density 
implied from the optical and MIR spectra appear to differ, it is 
interesting to ask whether we can construct a photoionization model 
which is consistent with {\it both} the optical and MIR line intensities. 
The photoionization model will concern only region A as both the optical 
and MIR slits include only that region. Radio observations 
of \citet{johnson04} show it to be composed of two main sources 
(A1 and A2) and a third much fainter one (A3),
separated by the angular distance $\sim$ 1\arcsec. 
Therefore, optical and MIR spectra in fact include the 
contribution of all three regions because of the relatively large apertures.
All three sources are thermal, indicating that the radio 
emission is from \hii\ regions. 

To construct a photoionized \hii\ region model of region A,
we have used the CLOUDY code \citep[version c05.07, ][]{cloudy,ferland98}.
We first ran a series of models assuming that both the optical and the MIR
light originate in the same region, characterized by the
Balmer-decrement derived extinction (\av\,=\,0.52\,mag).
While entirely consistent with the optical line emission,
these simple models underestimate the MIR line fluxes by factors of 5 or more.

We then developed a series of two-component models:
region A1 characterized by the
Balmer-decrement derived extinction (\av\,=\,0.52\,mag), and 
region A2, invisible in the optical.
The implicit assumption here is that the extinction \av\ in A2 is
$\ga$\,6 mag or so,
consistent with the absence of A2 in the \hst\
optical image \citep{johnson04}.
Hence, only A1 contributes to the optical emission, while
both A1 and A2 contribute to the emission in the MIR. 
We neglect the
contribution of region A3 as it is much less luminous than regions A1 and A2.
The dust responsible for the extinction is assumed to lie {\it outside} of
the \hii\ region, rather than mixed with the ionized gas.

The parameters of the \hii\ region model for A1 were adjusted such that it  
best reproduces the optical nebular emission-line intensities 
in \hh\ as observed by \citet{izotovthuan04}. 
Several input parameters are needed for this. The first is  
the observed flux of the \hb\ emission, corrected for the optically 
derived extinction \av\,=\,0.52 mag, of 4.59$\times$10$^{-13}$
erg s$^{-1}$ cm$^{-2}$. This corresponds to a \hb\
luminosity $L$(\hb) = 1.45$\times$10$^{40}$ erg s$^{-1}$, or a 
number of ionizing
photons \qlyc\,=\,3.05$\times$10$^{52}$ s$^{-1}$. The latter value is
consistent with that derived by \citet{johnson04} for region A1 
from the VLA radio continuum observations (after adjusting the distance). 
Another input parameter is the ionizing stellar radiation. 
We incorporate a Costar model \citep{SK97} with
a SED corresponding to that of a main-sequence star with an effective
temperature of 40600 K. We adopt a spherical ionization-bounded
\hii\ region model with an inner radius of 10$^{19}$ cm, a constant 
density of 100\,cm$^{-3}$, and a filling factor of 0.005. 
The chemical composition of the \hii\ region is shown in Table 
\ref{tab:chem_cloudy}. 
Only three parameters are fixed: \qlyc, the filling factor, and the gas number density. 
All other parameters were varied to obtain the best coincidence of the
observed and modeled line intensities in the optical range
(Table \ref{tab:optical_cloudy}). 
With this model for A1,
we obtain good or reasonable agreement for nearly all the optical line intensities. 
They correspond in general
to element abundance ratios typical of solar ones except for a 
slightly lower N/O abundance ratio. The exception is iron. To fit
the observed \feiii\ $\lambda$4658, $\lambda$5270 emission lines,
we needed to adopt a 
Fe/O abundance ratio which is about one order of magnitude lower
than the solar one. This suggests that iron is highly depleted in
\hh, probably onto dust grains, implying the presence of a significant
amount of dust in \hh. 
Our subsequent estimate of the dust mass ($\S$\ref{sec:dust}) confirms this 
conclusion. 

For region A2, we have no constraints in the optical range for
its H {\sc ii} region model. Therefore, we adopt for this region the same
parameters as for region A1 except for the number of ionizing photons \qlyc\ 
(i.e., \hb\ luminosity) and filling factor. According to \citet{johnson04}
\qlyc\ for A2 is 1.5 times greater than for region A1.
Thus, we assume that the intrinsic, i.e. extinction-corrected, 
flux of the \hb\ emission line in region A2 is 1.5 times greater than that in 
region A1.
As for the filling factor, we adopt it to be 0.0025 in region A2, 
lower than that for A1, 
because of the more diffuse morphology of A2 in the radio map by \citet{johnson04}.

The 05.07 version of CLOUDY includes the calculation of
line intensities in the MIR range. We combine both models for regions
A1 and A2 to fit the observed intensities of the MIR lines. 
In Table \ref{tab:mir_cloudy},
we compare the observed and predicted intensities of
the MIR lines listed in Table \ref{tab:lines}. Columns 2 and 5 
show the observed and predicted intensities for regions A1$+$A2
relative to the (intrinsic) total \hb\ intensity for both regions. 
Columns 3 and 4 give the predicted intensities of emission lines normalized 
to the respective intensities of \hb\ in each of regions A1 and A2. 
Column 6 is the ratio
of observed (column 2) to predicted (column 5) line intensity.
It is seen that the observed intensities of the majority of the MIR lines,
with the exception of the \neii $\lambda$12.81 $\mu$m and the
\neiii $\lambda$36.01 $\mu$m lines, 
are $\sim$ 1.7 times higher than the predicted ones. 

The Spitzer/IRS  aperture is significantly
larger than the narrow slit of 2\farcs0 used in the optical observations, so that
aperture effects could explain part of the discrepancy.
We can estimate the discrepancy factor potentially
due to aperture effects by using the total
observed \ha\ flux of 2.75$\times$10$^{-12}$ erg s$^{-1}$ cm$^{-2}$
measured by \citet{gildepaz03} on an \ha\ image of \hh. 
With the observed \ha/\hb\ ratio of $3.47$ in the optical slit,
and the uncorrected \hb\ flux of 2.6$\times$10$^{-12}$ erg s$^{-1}$ cm$^{-2}$,
we obtain an uncorrected (for extinction) \ha\ flux of
9.1$\times$10$^{-13}$ erg s$^{-1}$ cm$^{-2}$;
this value is about three times smaller than the total flux measured
by \citet{gildepaz03}.
We have also measured the fraction of the \ha\ emission in the \citet{gildepaz03}
image which falls in the IRS slits (see right panel of Fig.\ref{fig:slits}). 
Because of the orientation of the slits and their different
sizes, roughly 36\% 
of the total \ha\ emission lies within the SL slit; $\approx$25\% 
within the SH slit; and $\approx$80\% within LH.
Diffraction effects have not been considered here, so such estimates
are rather crude.
Moreover, the optical line emission suffers from highly variable extinction,
which would result in relatively less \ha\ flux in the more obscured region A2.
All things considered, we conclude that
aperture effects could account for the 
discrepancies for most lines in Table \ref{tab:mir_cloudy}. 

The discrepancy between the observed and predicted intensity ratios of the
two weakest lines, \neii $\lambda$36.01 $\mu$m and
\neii $\lambda$12.81 $\mu$m, are larger than for the others.
In part, this discrepancy could be explained by the observational errors, at
least in the case of the weakest \neiii $\lambda$36.01 $\mu$m emission line.
Indeed, the observed ratio \neiii $\lambda$15.55 $\mu$m /
\neiii $\lambda$36.01 $\mu$m of $\sim$ 6 is a factor of two
greater than the predicted one.
As for the \neii $\lambda$12.81 $\mu$m emission line, 
its intensity was obtained with a
deblending algorithm from the high-resolution IRS SH spectrum (see $\S$\ref{sec:lines})
because of its proximity with the 12.7\,\micron\ PAH feature.
Even with the high-resolution IRS spectrum, this deblending is not straightforward,
and the \neii\ line could consequently be overestimated. 

Thus, taking into account a high extinction for region A2, 
aperture corrections and possible sources of
the uncertainties for the weakest MIR lines, we find an overall good agreement
between the observed intensities of the optical and MIR lines.
We conclude that the MIR lines probe a region 
which has more extinction than the location where the optical lines 
originate. This hypothesis will be confirmed later by our \dusty\ 
calculations.

\section{\label{sec:morph}The Infrared Morphology of \hh}
 
\begin{figure}
\vspace{0.2cm}
\hbox{\includegraphics[angle=0,bb=87 49 566 607,width=0.4\linewidth]{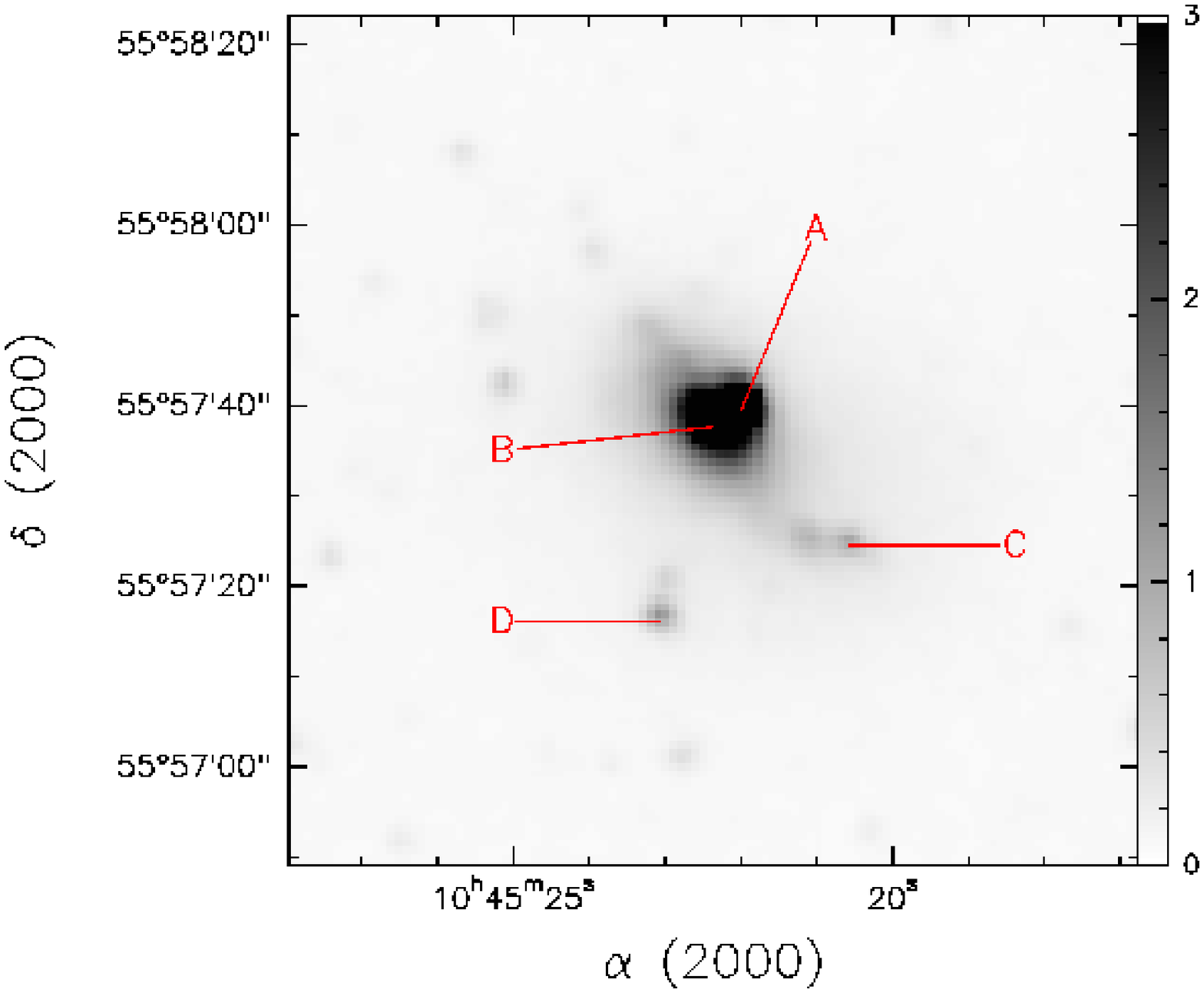}
\hspace{1.5cm}\includegraphics[angle=0,bb=87 49 566 607,width=0.4\linewidth]{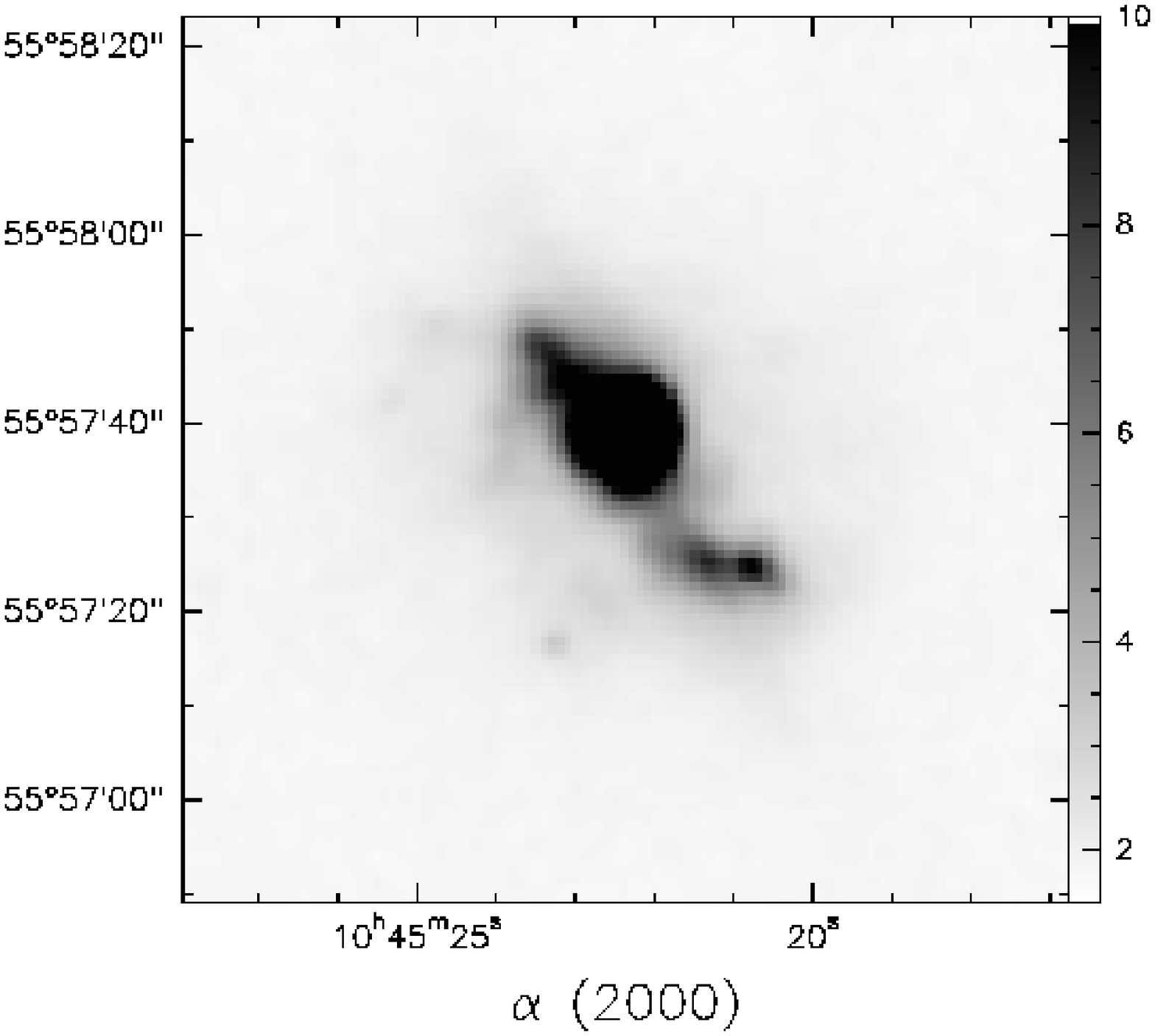} }
\caption{The 4.5\micron\ (IRAC2) image (left panel) and
the 8.0\micron\ (IRAC4) image (right panel) of \hh.
The units given in the colorbar are MJy/sr.
The star-forming knots identified by \citet{steel96} are labeled.
\label{fig:irac}}
\end{figure}

We now use the Spitzer IRAC 
(Fig. \ref{fig:irac}) and MIPS (Fig. \ref{fig:mipsoverlay}) data 
to study the origin of the infrared emission in \hh. 
Fig. \ref{fig:irac} shows that both the 4.5\,\micron\ (IRAC2) and 8.0\,\micron\ (IRAC4) 
emissions are extended on galaxy scales (see also the growth curves of 
as compared to the point-spread function in Fig. \ref{fig:phot}). 
They follow well the optical features as shown in 
Fig. \ref{fig:iracoverlay}, and the optical knots A$-$D are all detected. 
In particular, the IRAC emission peaks on region A, 
coincident with the strong compact radio source seen by \citet{johnson04}, 
the site of the most active star-forming region in \hh. 

\subsection{The PAH Component}

Emission in the IRAC bands is the sum of 
three different components: 1) the stellar 
photospheres, 2) the hot dust continuum, and 3) the PAHs. 
Because of the usual dominance of the PAHs in MIR spectra
of star-forming galaxies (see the \hh\ MIR spectrum in Fig. \ref{fig:irs}), 
the 8\,\micron\ IRAC channel is commonly used as
a ``PAH indicator'' \citep{hogg05,engelbracht05}.
 The 4.5/8.0 flux ratio is thus a potentially important diagnostic
for characterizing the ISM properties at short wavelengths:
high 4.5/8.0 ratios would indicate predominantly stellar
light, whereas low ratios would point to a relatively large PAH
contribution.

\begin{figure}
\vspace{0.2cm}
\centerline{ \includegraphics[angle=270,width=0.52\linewidth]{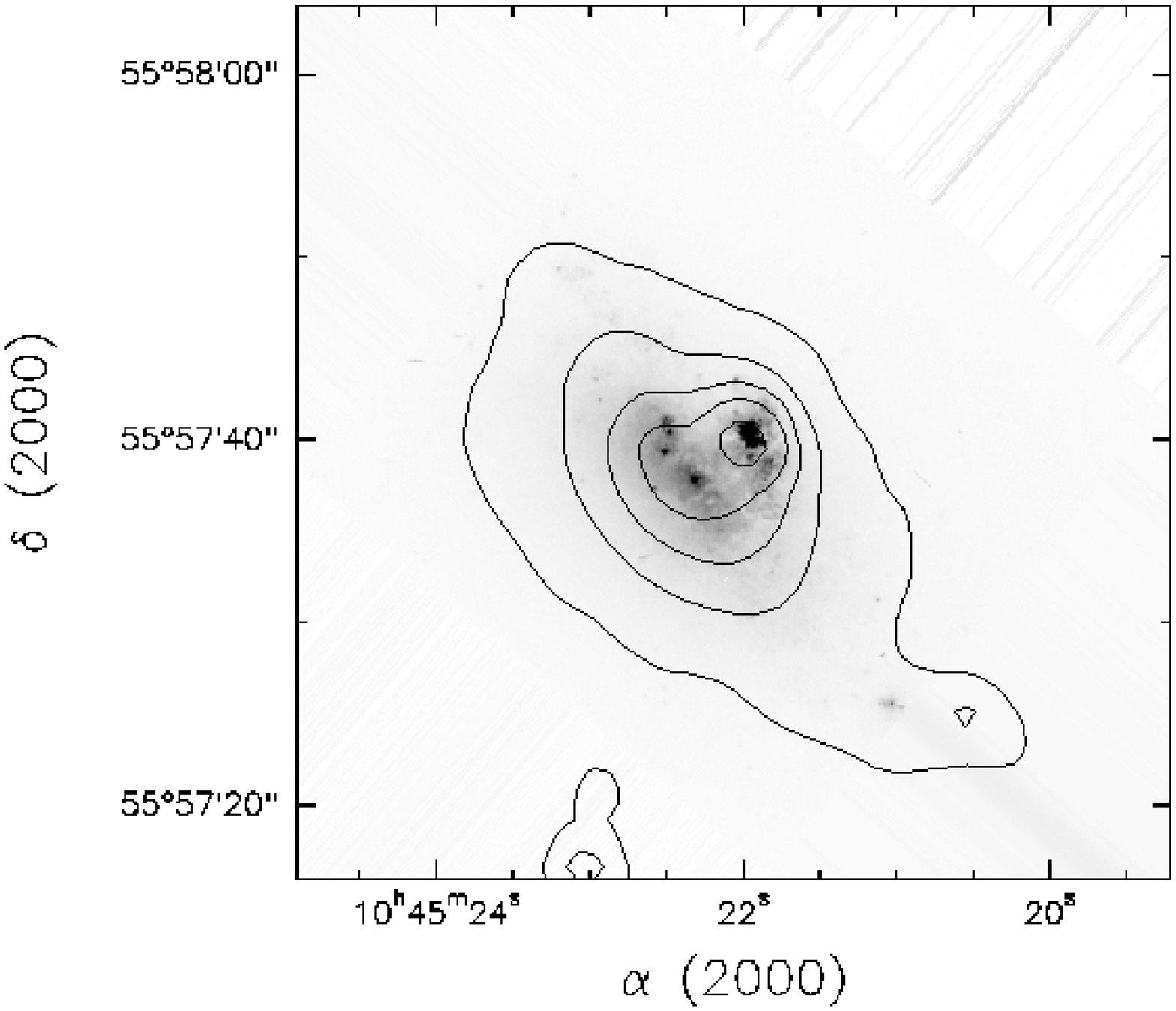} 
\hspace{0.5cm}\includegraphics[angle=270,width=0.5\linewidth]{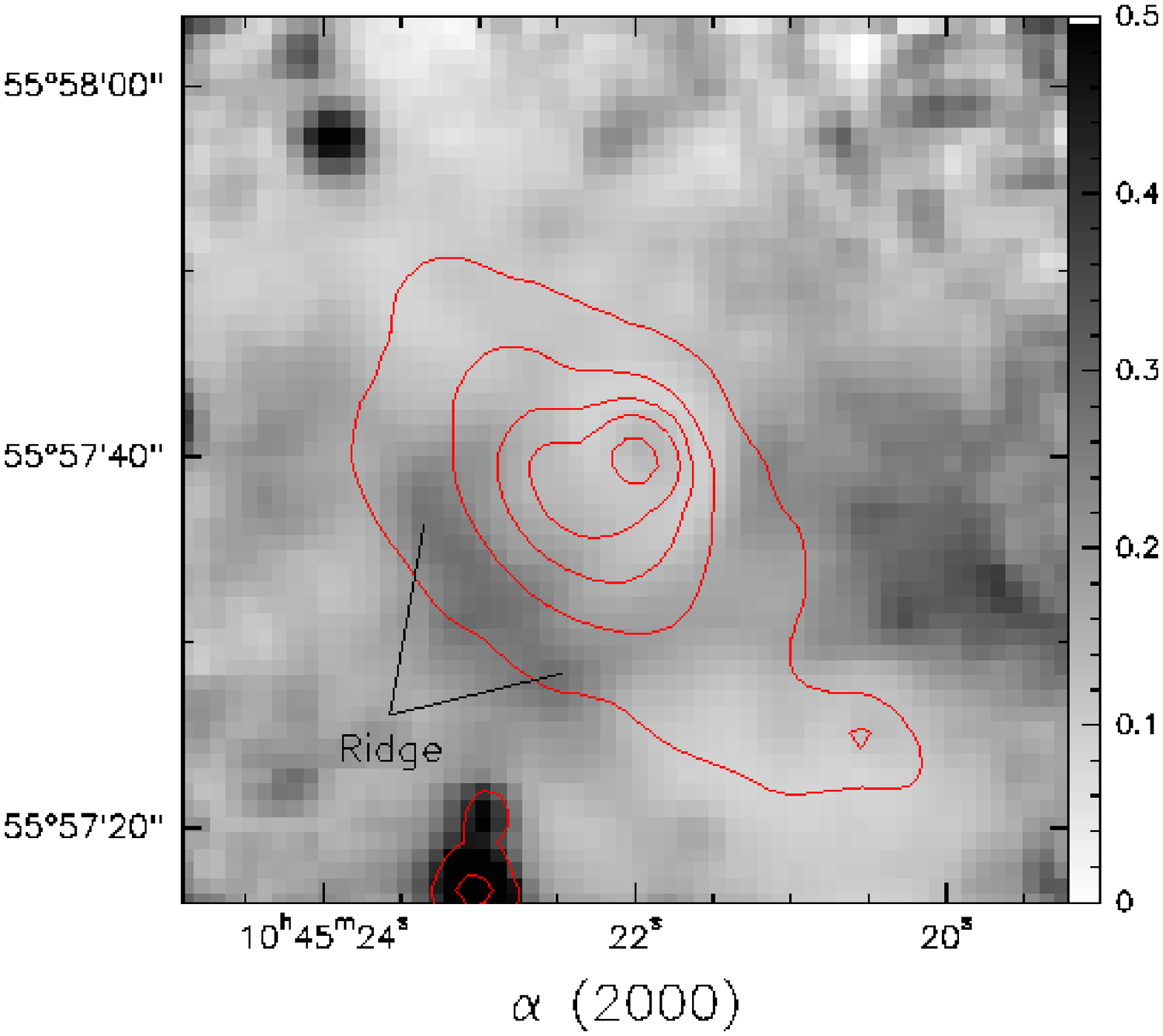} }
\caption{The 4.5\micron\ IRAC image of \hh\ (see Fig. \ref{fig:irac}) 
contoured on the (left panel) \hst /WFPC2 F606W image \citep{mgt}
and on the 4.5/8.0\micron\ flux ratio (right panel).
Contours range from 2, 5, 10, 20, 45, 70, 95\% of the 4.5\,\micron\ peak;
the ``ridge'' feature referred to in the text is marked.
The bright optical point source is clearly coincident with the 4.5\micron\
brightness peak.
The 4.5/8.0\micron\ ratio changes by a factor of 10 over the image. 
\label{fig:iracoverlay}}
\end{figure}

An IRAC 4.5/8.0 color image was created from the two IRAC 4.5 and 8\micron\ 
mosaics
by first subtracting the background measured in empty regions in each mosaic, 
then aligning the two mosaics by cross correlation, and finally
dividing one by the other.
The result is shown in the right panel of 
Fig. \ref{fig:iracoverlay} where the 4.5\micron\ 
image contours are superimposed on  the 4.5/8.0\,\micron\ flux ratio map.
Inspection of Fig. \ref{fig:iracoverlay} shows that the 4.5/8 flux
ratio varies by more than a factor of 10 over the galaxy, with peaks
$\ga$0.5 and troughs of $\la$0.1.
Because the 4.5/8.0 color of stellar populations is $\sim2.5$,
roughly independently of age and metallicity \citep{sb99},
it is clear the the color map of \hh\ is dominated by the ISM,
and not by stars. 
The only source with 4.5/8.0 colors near the stellar value 
is knot D to the south of the nucleus (there is no \ha\ 
emission associated with this object).

If we interpret the 8\,\micron\ emission in the IRAC band in the 
``standard'' way and attribute it primarily to PAHs, 
the low 4.5/8 ratio $\sim\,$0.1 around knots A and B (see Fig. 
\ref{fig:iracoverlay}) would imply a substantial PAH contribution. 
Indeed this interpretation is supported by the detection of
strong PAH features in the IRS spectrum (see $\S$\ref{sec:spectra}).
If all the changes in the 4.5/8 ratio over the galaxy were attributed only to
changes in the PAH emission, the higher ratio $\sim\,$0.3 in the broad ``ridge''
to the southeast of the nucleus (see Fig. \ref{fig:iracoverlay})
would indicate that PAHs should be {\it three times less prominent} there.
It could be that the PAHs are destroyed by the radiation fields in the
star-forming regions associated with the intense circumnuclear \ha\ visible
in the right panel of Fig. \ref{fig:slits}.

\subsection{The Hot Dust Continuum}

On the other hand,
the spatial variations in the 4.5/8 flux ratio could be
associated with variations in the hot dust 4.5$-$8\,\micron\ continuum.
As pointed out by \citet{wu06} and others, 
variations in the 8\,\micron\ continuum can
masquerade in broadband images as variations in the PAH contribution.
Because in \hh\ the PAHs and continuum contribute roughly equally to 
the 8\,\micron\ filter\footnote{This is indeed the case for region A: 
we decomposed the IRS spectrum into a sum of lorentzian functions for 
the PAH bands and a continuum and synthesized from this model the 
IRAC 8\,\micron\ flux, to show that each component contributes the same 
amount of energy in the band.},
it is not unreasonable to think that the larger 4.5/8\,\micron\ ratios
are due to a flatter continuum slope, i.e. a lower 8\,\micron\ continuum,   
in the general extended ISM as compared to star-forming region A.

\begin{figure}
\vspace{0.2cm}
\centerline{ \includegraphics[angle=270,width=0.5\linewidth]{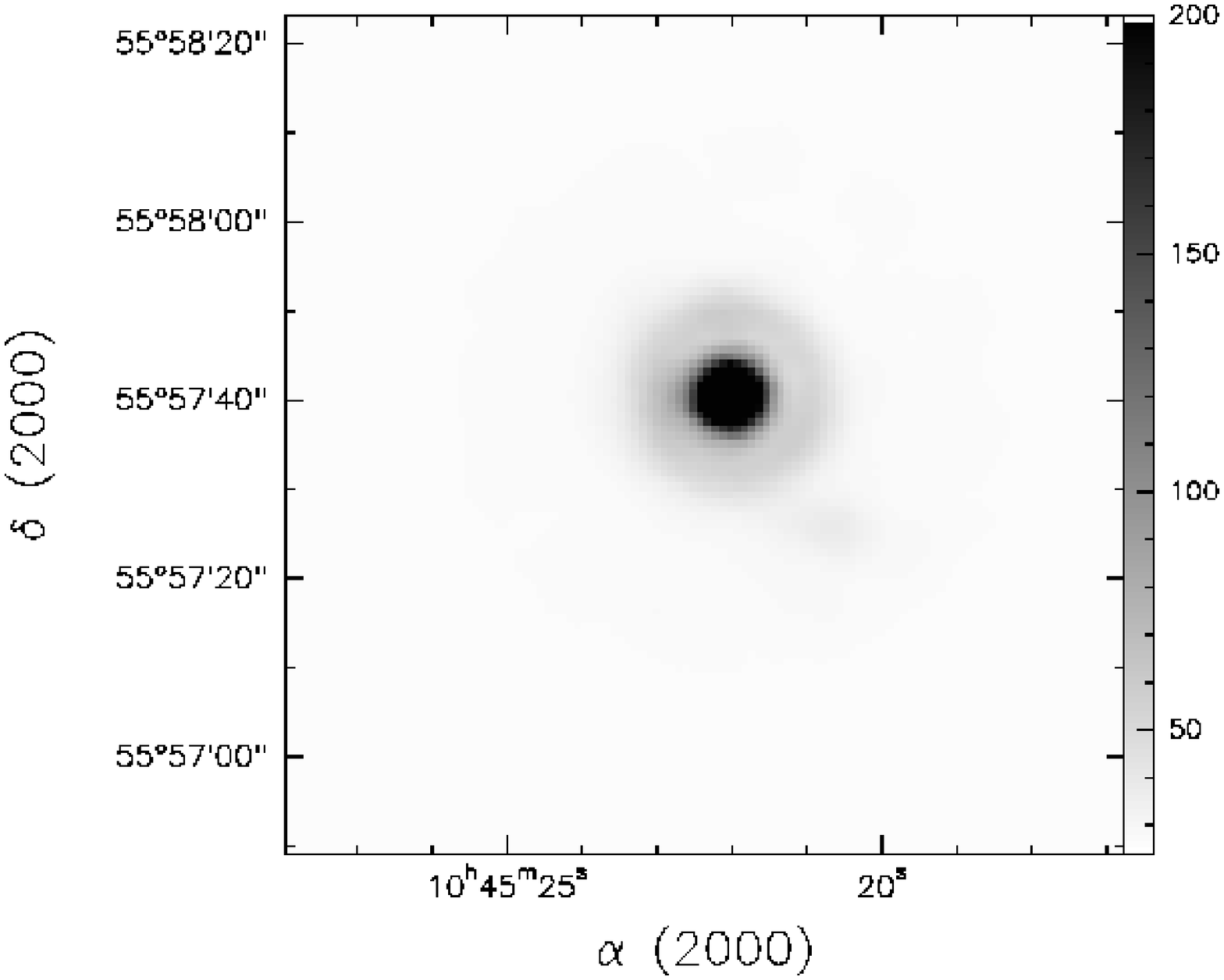} 
\includegraphics[angle=270,width=0.5\linewidth]{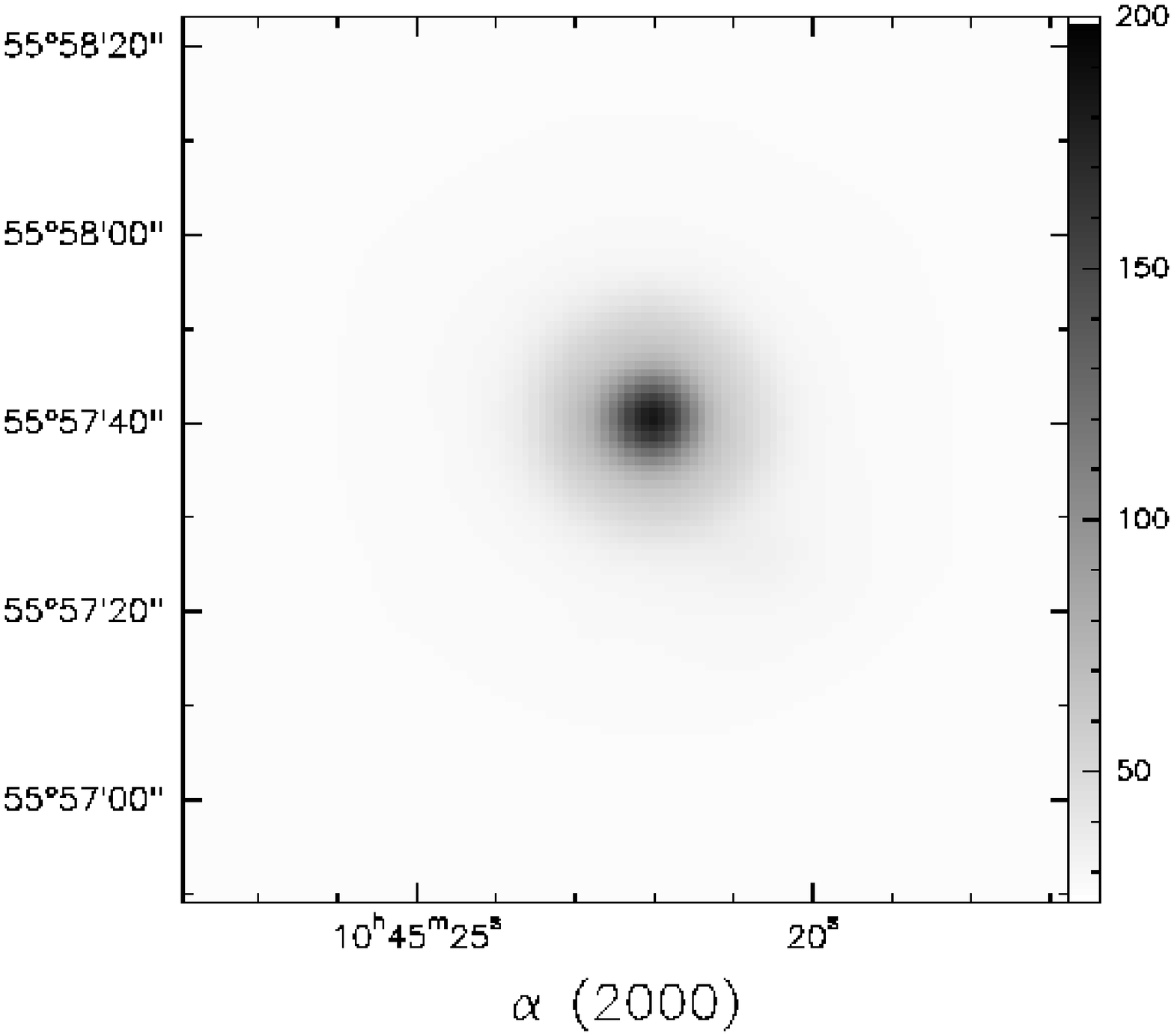}}
\caption{The 24\,\micron\ mosaic of \hh\ before (left panel) and after (right
panel) deconvolution.
Units of the colorbar are MJy/sr.
\label{fig:mips24}}
\end{figure}

\begin{figure}
\vspace{0.2cm}
\centerline{ \includegraphics[angle=270,width=0.50\linewidth]{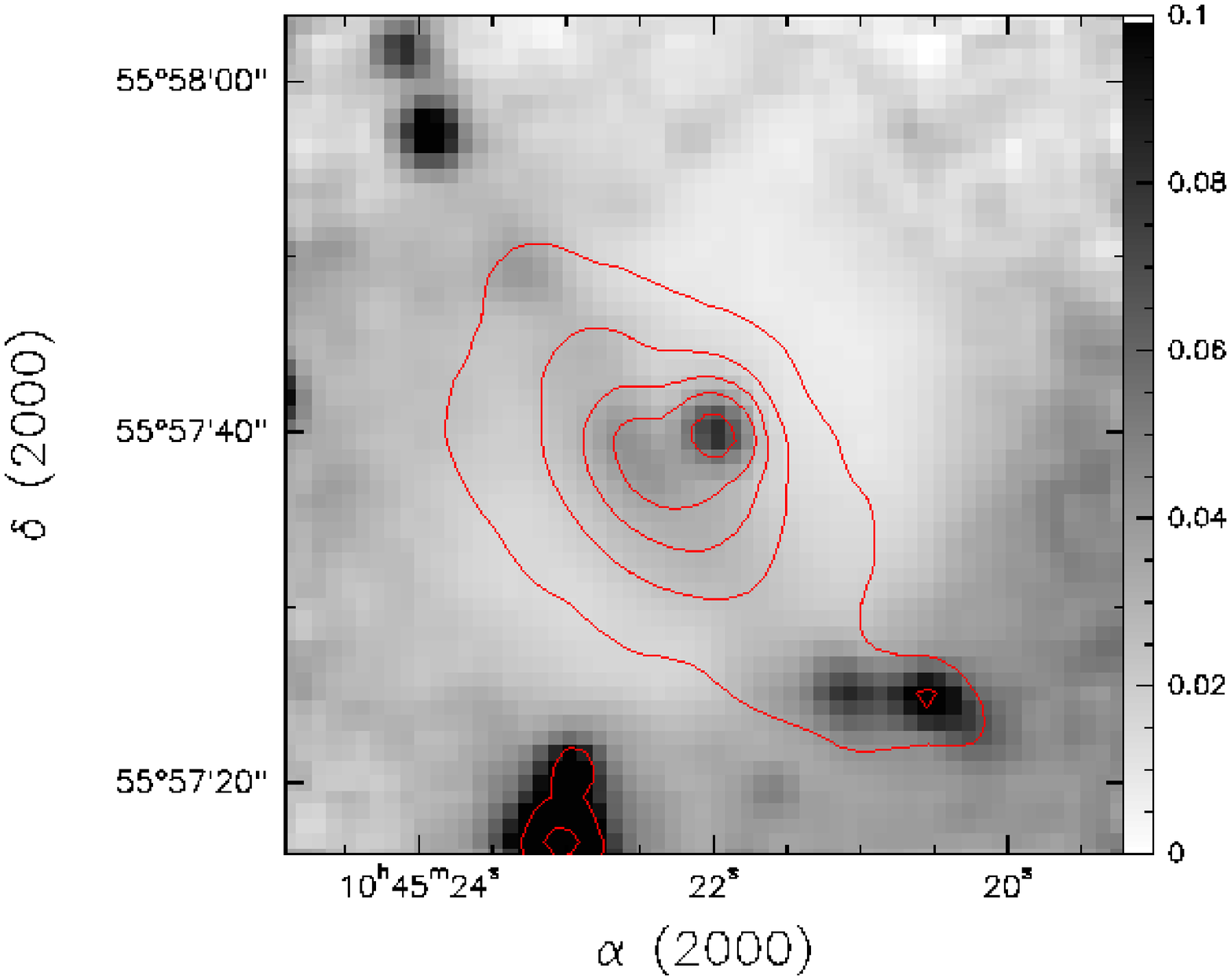} 
\hspace{0.02\linewidth} \includegraphics[angle=270,width=0.45\linewidth]{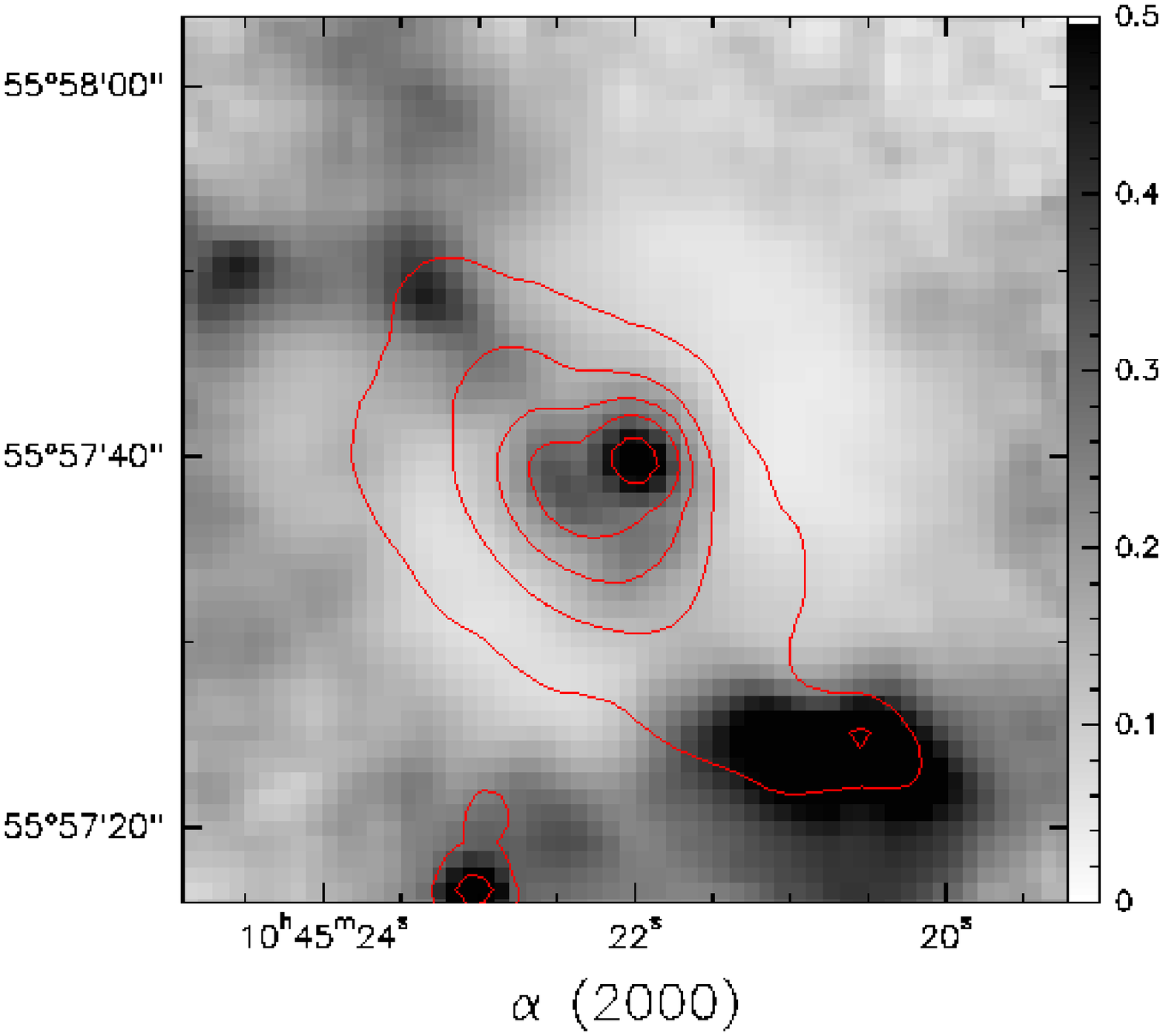}}
\caption{Contours of the 4.5\,\micron\
image superimposed on the 4/24\,\micron\ flux ratio (left panel), and on
the 8/24\,\micron\ flux ratio (right panel).
The 24\,\micron\ image has been deconvolved.
\label{fig:iracmips}}
\end{figure}

We can test this hypothesis by making IRAC/MIPS hybrid color images.
Since the IRAC 4.5\,\micron\ image traces stars and hot dust but not PAH 
features, 
the IRAC 8\,\micron\ image mainly the dust continuum and PAHs 
with a very small stellar contribution, and
the MIPS 24\,\micron\ image the warm or hot dust continuum only, 
we should be able to use the IRAC/MIPS colors to distinguish between 
continuum and PAH variations.
Since the Spitzer 
diffraction limit is 1\farcs11 at 4.5\,\micron, 
1\farcs98 at 8\,\micron\ (see the IRAC Data Handbook),
and $\sim$6\arcsec at 24\,\micron,
comparison of the IRAC 4 and 8\,\micron\ images
 with the MIPS 24\,\micron\ image
is only possible after some processing of the latter.
The availability of an accurate PRF makes possible
a deconvolution for the MIPS-24 image, 
which we performed with the {\it lucy} task in the
{\it STSDAS\,}\footnote{{\it STSDAS} is the
Space Telescope Science Data Analysis System,
distributed by the Space
Telescope Science Institute, which is operated by the Association of
Universities for Research in Astronomy (AURA), Inc., under NASA contract
NAS 5--26555.}
extension of {\it IRAF},
using the {\it MIPS24\_PRF\_mosaic} 
in the {\it mopex} distribution as the deconvolution kernel.
Prior to deconvolution,
the MIPS image was linearly interpolated
to the finer pixel scale of the PRF.
After deconvolution, the image was rebinned to the original scale
of the MIPS\,24\,\micron\ mosaic (see $\S$\ref{sec:mipsdata}), 
designed to be the same as the IRAC scale.
The original and deconvolved MIPS-24 images are shown in Fig. \ref{fig:mips24}.
Just as we did for the IRAC color maps, background was first subtracted 
for the IRAC and MIPS images, then 
the images were aligned using a cross correlation technique,
and finally the IRAC image was divided by the MIPS image. 
The 4/24\,\micron\ and 8/24\,\micron\ flux ratios obtained from the
deconvolved 24\,\micron\ images are shown in 
Fig. \ref{fig:iracmips} where the 4.5\micron\ 
image is superimposed in contours.

There are several features of the hybrid IRAC/MIPS-24 color images in 
Fig. \ref{fig:iracmips} to be noticed.
First, the high ratio of $\sim$0.6 in source A in the
8/24 color map is much larger than
the value inferred from the IRS spectrum of $\sim$0.06.
However the spectrum has not been deconvolved and is roughly
diffraction limited at all wavelengths.
This means that it encompasses a 
larger fraction of the extended source as wavelength increases. 
At $\sim$8\,\micron\ the spectrum encompasses a region 
of $\sim$2\arcsec\ in size,
while at 24\,\micron\ it encompasses a $\sim$6\arcsec\ region.
Therefore, it is likely that the higher image value is closer
to the true one for region A than the low value derived from the IRS spectrum
which is more typical of the general ISM.
Indeed, the value of the 8/24 color averaged over a 40\arcsec\ square
region is $\sim$0.17, consistent with the color derived from total fluxes.
Hence, we believe that the general structure of the hybrid IRAC/MIPS-24 
color images is reliable.

\begin{figure}
\vspace{0.2cm}
\centerline{ \includegraphics[angle=270,width=0.53\linewidth]{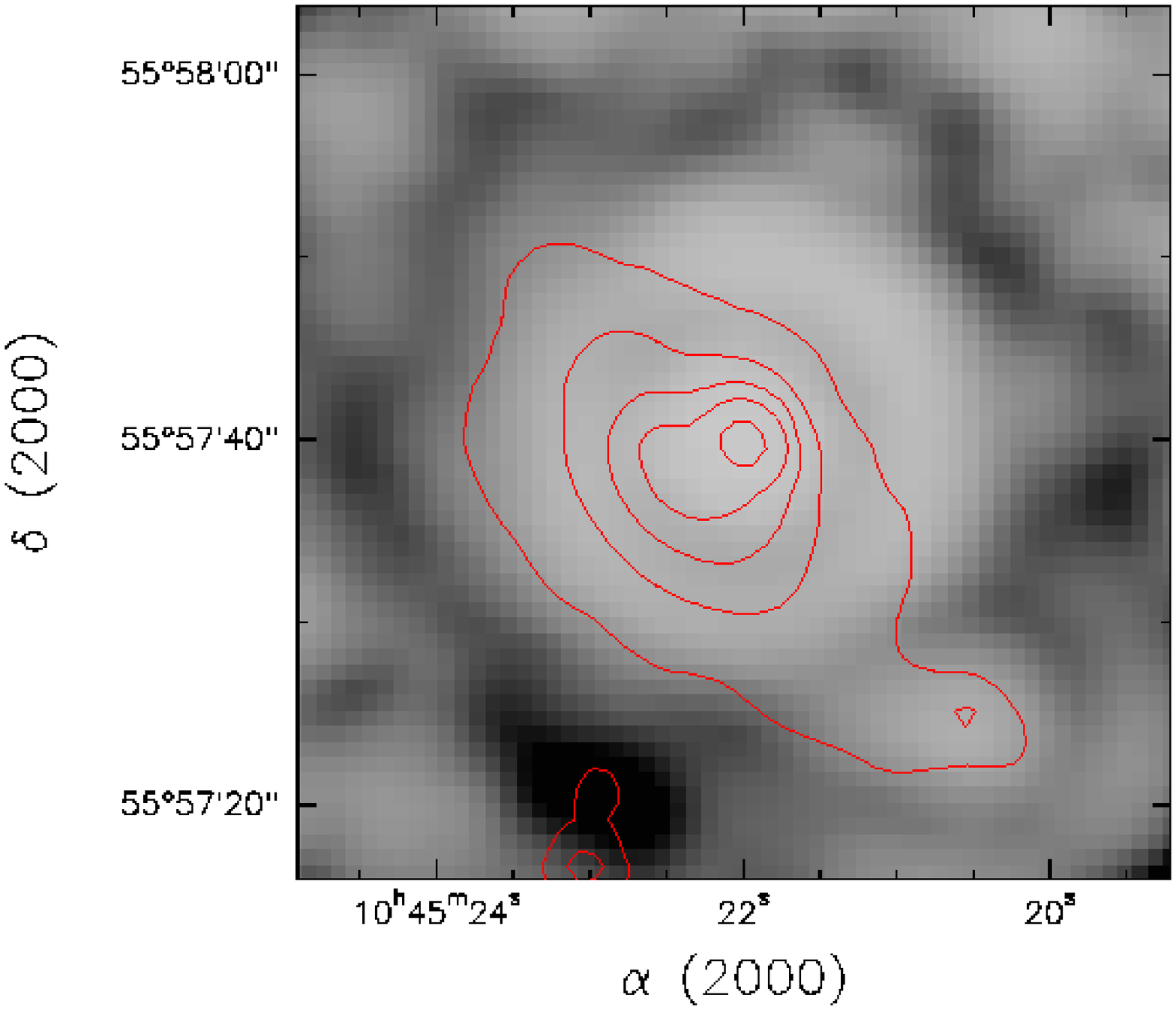} 
\includegraphics[angle=270,width=0.48\linewidth]{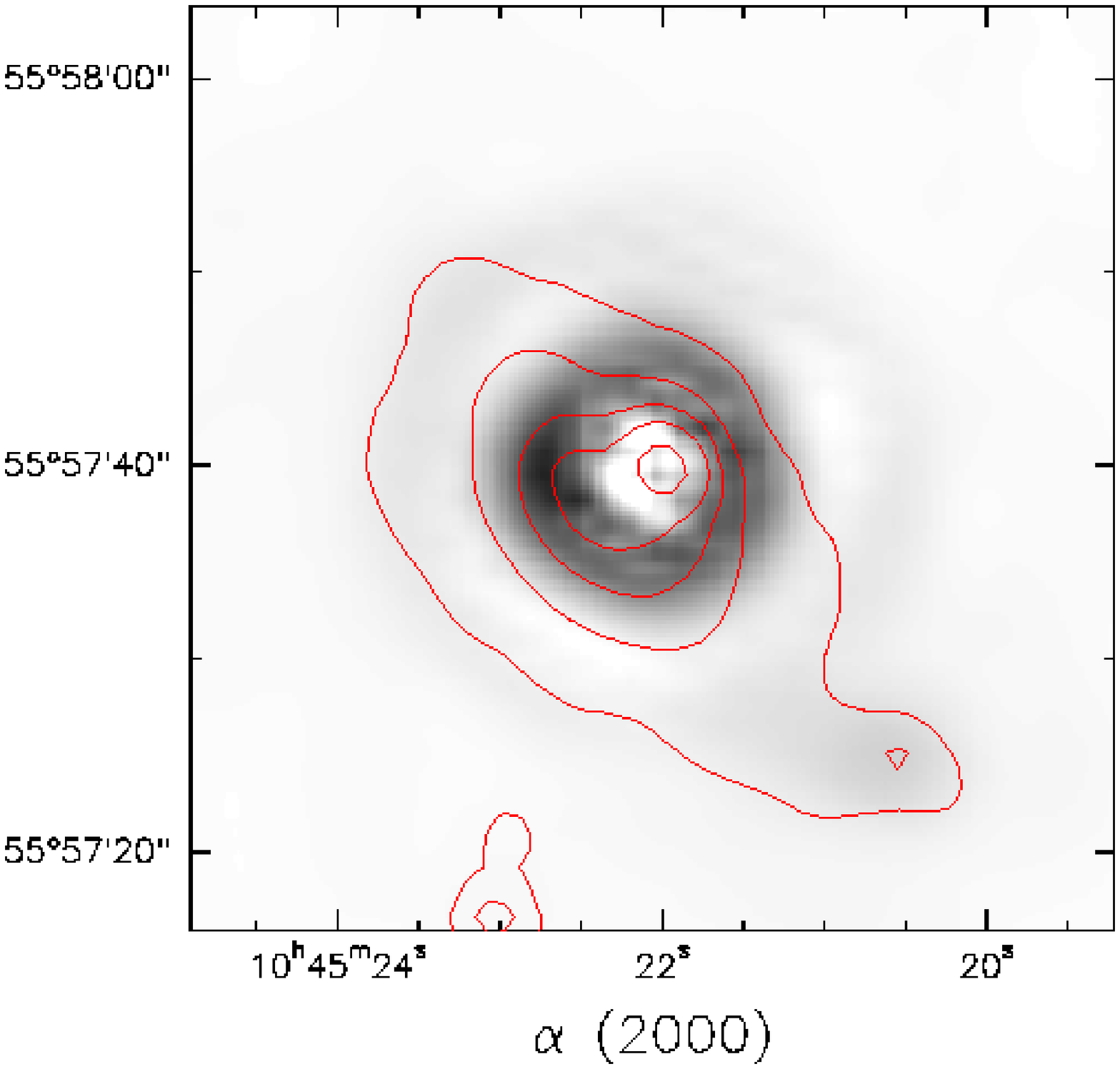}}
\caption{Results of two manipulations of the MIPS-24 image.
The left panel shows
the 8\,\micron\ image convolved with the MIPS-24 PRF, then divided by the original
24\,\micron\ image; 
the right panel gives the 24\,\micron\ image obtained by subtracting the MIPS-24 PRF
from the original 24\,\micron\ image.
In both panels,
the IRAC 4.5\,\micron\ is overlaid in contours as in Fig. \ref{fig:iracoverlay};
intensity grows from white to black.
\label{fig:mipsops}}
\end{figure}

Second, as mentioned before,
there is a noticeable feature with a high 4.5/8 ratio ($\sim\,$0.3) 
tracing a broad ridge to the southeast of region A 
(see the right panel of Fig. \ref{fig:iracoverlay}).
Our aim is to use the 24\,\micron\ map to trace the hot dust continuum,
and compare it to the IRAC colors.
The ridge feature is also present in the 4.5/24 and 8/24 color maps 
(Fig. \ref{fig:iracmips}).
However, because of its location, the feature in the 24\,\micron\ map
ratios may be an artefact of the deconvolution procedure.
To check its reality, we have performed three tests.
First, we have convolved 
the IRAC 8\,\micron\ image with the MIPS-24 PRF and then divided the 
resulting image 
by the original 24\,\micron\ image. Second, we have 
convolved the MIPS-24 deconvolved image with the IRAC 8\,\micron\ PRF and 
divided by the original 8\,\micron\ image. Third, 
we have divided the original 24\,\micron\ image by its PRF. 
The results of the first and third tests are shown in Fig. \ref{fig:mipsops}. 
The second test is not shown:  
because of the small size of the IRAC PRF,
the color image obtained this way is virtually indistinguishable 
from the one made with the MIPS-24 deconvolved image.
Inspection of Fig. \ref{fig:mipsops} shows that 
the feature of interest in the 24\,\micron\ map coincides
roughly with the Airy wings of the MIPS-24 PRF. Thus the broad ridge 
in the 24\,\micron\ map ratios may not be a real feature; 
this means that the 24\,\micron\ image cannot be used to 
unequivocally validate either
hypothesis for the red 4.5/8\,\micron\ ridge color.

Nevertheless, we can examine the relative colors of the ridge and the
surrounding disk.
First, the {\it ridge} in \hh\ has a 4.5/8 flux ratio of $\sim0.28$, 
``bluer'' than the surrounding {\it disk} with $\sim0.12-0.15$.
If the change in color depends solely on the 8\,\micron\ band,
then the ridge
would have to have either 50\% fewer PAHs or 50\% less hot dust
at 8\,\micron\ than the disk.
Since we expect the disk to have less hot dust than any actively star-forming
region, we must necessarily conclude that the ``ridge'' is deficit in PAHs,
not in the hot dust continuum.
The PAHs could have been destroyed because of proximity to the edge of the ionized gas shell
(see \ha\ image in Fig. \ref{fig:slits}), or by supernovae shocks
and outflows perpendicular to the dust lane.
        
The morphology of the 4/24 and 8/24 flux ratio maps as shown in 
Fig. \ref{fig:iracmips} is complex; 
both ratios vary by a factor of $\ga$30 over the galaxy, being greatest 
in and around knot C toward the southwest and knot D toward the southeast.
Because knot D has {\it R}-band emission, but no \ha\ in the images by \citet{gildepaz03},
the reason for its high 4/24 and 8/24 ratios is unclear; 
indeed we cannot exclude that knot D be merely a foreground star.
On the other hand, in knot C, 
we tentatively interpret these high ratios to be due to {\it increased}
PAH strength, or equivalently, {\it decreased} hot dust. 
First, the 4.5/8 flux ratio in knot C is half that in knots A and B, 
and the morphology in the color is clearly extended.
Second, both the 4.5/24 and 8/24 ratios are also lower in knot C. 
Hence, we conclude that there 
is less hot dust emitting at 4.5\,\micron\ and 8\,\micron\ in knot C than in 
knots A and B, and possibly an 8\,\micron\ PAH excess. 

The difficulty of separating PAH and continuum contributions in the broadband
IRAC 8\,\micron\ filter is evident from the above discussion.
In \hh, with its conspicuous \ha\ emission throughout the entire central region
and across the dust lane, the ISM changes from predominantly neutral in the
outer disk to ionized in the inner regions.
Because of the complex morphology from 5 to 8\,\micron\ on kpc scales,
we conclude that the ISM has been disturbed by the combined effects of the
star-formation episodes concentrated in knots A, B, and C. 

\subsection{The Cool Dust Continuum}

Fig. \ref{fig:mipsoverlay} shows the dust emission at the longer wavelengths of 24, 70 
and 160\,\micron. 
The emission at 24 and 70\,\micron\ is virtually point-like, and appears to be mainly 
associated with the most active star-forming region A. 
The point-like nature of the 
24 and 70\,\micron\ emission is also illustrated by the growth curves in Fig. \ref{fig:phot}
\footnote{The 24\micron\ flux given by the gaussian fit 
discussed in $\S$\ref{sec:phot} is too low relative to the spectrum.
The MIPS-24 image of \hh\ resembles the octagonal shape of the PRF, which
is strongly diffraction limited with a pronounced Airy ring. 
Hence, we would expect a gaussian fit to give a lower flux than the true one, 
since it does not include the flux in the Airy maximum at radii of $\sim$6\arcsec,
and may also have overestimated the background because of the additional emission at these radii.}. 
However, the emission at 160\,\micron\ appears to be more extended (see also its 
growth curve in Fig. \ref{fig:phot}). 

Like most other galaxies,
\hh\ is a composite entity in the IR; we see different 
components and morphologies depending on wavelength. In \hh,
we see {\it extended} stellar photospheric and {\it compact}
hot dust continuum emission at 4.5\,\micron; 
{\it extended} PAH emission in the general ISM at 8\,\micron\ 
but with some hot dust continuum and stars;
{\it compact} warm dust emission associated with 
active star-forming regions at 24 and 70\,\micron; and cooler {\it extended} dust 
emission associated with older stellar populations at 160\,\micron.  

\section{\label{sec:dust}Modeling the 
Infrared Spectral Energy Distribution of Region A in \hh}

We now use the complete IR SED to constrain the 
properties of the star formation and the dust in the brightest 
star-forming region in \hh, region A. 
The infrared morphology revealed by the IRAC and MIPS images 
(Fig. \ref{fig:irac} and \ref{fig:mipsoverlay}) 
suggest that region~A dominates the SED of \hh\
at wavelengths $20\,\la \lambda \la\,100$\,\micron.
The IR SED of this region, a
strong continuum with ionic lines indicating a relatively hard
ionizing source, is quite similar to that observed toward
dust-enshrouded super star clusters, such as for instance in 
NGC\,5253 \citep[see ][and references within]{van04}. Such an identification is
confirmed by the detection of compact thermal radio emission from
region~A by \citet{johnson04}, emission identified as originating in
the compact \hii\ region created by the star cluster.

\subsection{Fitting with \dusty}

To constrain the properties of this cluster and its surrounding dust
envelope, we have fit the NIR-FIR SED of this source with the
radiative transfer code \dusty\ \citep{dusty,elitzur}.
\dusty\ solves the radiative transfer problem in a dust envelope but has
a number of limitations which must be understood so we can properly evaluate 
the results it gives. First, \dusty\ assumes a spherically
symmetric dust envelope which is decoupled from the ionizing 
radiation source. While this decoupling is probably realistic for the stars and
the dust (we do not expect much of the dust to survive in or
very near the star cluster), it is much less so for the ionized gas 
which is a source of radiation through its continuum emission, and
the dust. The assumption of spherical symmetry is also too strong to be
realistic. Second, \dusty\ assumes that grains are in thermal equilibrium
with the radiation field. This may be the case for all grains near and in 
the cluster exposed to its intense and hard radiation, but clearly not so 
for grains in the general interstellar medium of \hh. These two
limitations mean that the SED we can fit realistically with \dusty\ concerns only
region~A, but not the whole galaxy.

To construct the SED of region~A, we must use high spatial resolution
observations in order to distinguish its emission 
from emission from other parts of \hh. In the NIR, which is a useful domain to constrain
the total optical depth to the cluster, we use the NIR 
observations listed in Table \ref{tab:photom}, taken from \citet{johnson04}. 
Given that our IRS spectrum is centered on region~A and includes the spectral region 
covered by our IRAC data, we have preferred to use the IRS rather than the IRAC 
data in the 5-35\,\micron\ range. 
We however do not use the IRS spectrum in its original form 
but rather in its PAH-subtracted version. As explained earlier, the PAH
component in the original IRS spectrum is related to the general ISM of \hh\
and therefore not relevant here. The MIPS point at 24\,\micron\ is
included in the fit, although it does not provide supplementary
constraints with respect to the IRS spectrum. As the 70\,\mic\ MIPS
image is strikingly pointlike we assume that it is entirely due to
region~A. The 160\,\micron\ MIPS image is more extended than the PSF
(Fig.~\ref{fig:phot}) and therefore likely contains a contribution
from the general ISM. However it is not possible to accurately
deconvolve the 160\,\mic\ image to derive the point-source component flux
only. We thus should keep in mind in judging the goodness of the fit
that the 160\,\micron\ flux is likely an upper limit to the actual flux
from region~A. 

The IRAS and ISO measurements of \hh\ were not used in
the fit for two reasons. First, the ISO fluxes and the IRAS 12 and
25\,\mic\ fluxes fall in a region of the SED that is better covered by
the IRS and MIPS data, and including them does not add more information. 
Comparison of the IRAS 60 and 100\,\micron\ fluxes to the MIPS 70 and
160\,\micron\ fluxes shows that, although they are compatible with each
other within their respective error bars, the IRAS photometry is
systematically higher than the MIPS photometry. This can be due either
to the larger IRAS beam since it subtends more faint extended
emission than the smaller Spitzer beam or to an incorrect
cross-calibration between IRAS and Spitzer. At this stage of the
Spitzer mission, cross-calibration with another mission can indeed be
uncertain and, as a result we rather restrict ourselves to the Spitzer
data. Finally, the radio data of \citet{johnson04}, although of very
high spatial resolution, were not used in the fit because 
first, \dusty\ does not take
into account radio emitting processes; and second, the contribution of
free-free and synchrotron emission to the SED at 
the longest wavelength point (160\,\micron) is negligible.

Before proceeding with the fit, we recall that region~A is 
composite. It is 
composed of two main sources (A1 and A2) and a much fainter third one
(A3). 
Making a single \dusty\ model for a collection of identical
sources is not a problem since the total values of the quantities of interest
such as opacities, luminosities and masses, are independent of the actual
number of sources used to fit the SED. 
However, as discussed before, 
A1 and A2 are not identical (we will neglect the faint A3). 
Thus, A2 has no optical counterpart in the \hst\/F606W band image while A1 does.
It is also not clear that the Br$\gamma$ and radio
emission in the direction of A2 come from the same source, while such an 
association is much clearer in A1. Finally, the different derivations
of the extinction toward region A (essentially A1+A2) by 
\citet{johnson04}, using the radio to Br$\gamma$ ratio, the
hydrogen Balmer line ratios, and the NIR colors, disagree. While this is
a rather common situation, it points to a non-homogeneous extinction
in region A and suggests that A1 and A2 may not be in the same
``evolutionary'' stage. 
Indeed, our CLOUDY modeling does suggest that A2 has more extinction than A1
($\S$\ref{sec:cloudy}).
This composite nature of region A must be kept in mind when
discussing the physical properties deduced from the fit. This is however the 
best we can do: the 
lack of high resolution data in the MIR-FIR allows us only to make 
a single \dusty\ model of the SED of region A which includes both A1 and A2.

When judging the goodness of the fit, it is also important to compare
the number of independent constraints with the number of model
parameters. For practical purposes we have integrated the IRS data in a
series of 10 pseudo-filters with a resolving power $\lambda/\Delta\lambda$ 
of 5. The NIR data
consists of three bands, while MIPS provides 2 independent
measurements. We therefore have 15 independent constraints on the
fit. The model requires a large number of parameters. We need first to 
specify a heating source. We take it to be described by the
Starburst99 \citep[][ hereafter SB99]{sb99} burst stellar populations at the
metallicity of \hh. This requires one parameter, the age of
the burst. The dust is described by its composition and size
distribution. We take it to be formed of silicates, graphites and
amorphous carbons. Thus the dust mix requires two more parameters. The dust
size distribution is characterized by a minimum and a maximum size and 
a power law exponent, requiring three additional parameters. 
The dust shell around the star cluster
is modeled as a series of embedded shells, each with an outer (relative) radius 
and an exponent for the power-law describing the dependence of the 
dust density on radius. Thus
the total number of parameters required is $2 \times N_{\rm
shell}$. The ``size'' of the system is described by the temperature at
the inner edge of the first shell. Finally the total amount of dust is
constrained by the optical depth at V. A complete \dusty\ model is
therefore described by $8 + 2 \times N_{\rm shell}$, i.e. in
principle, a model with 3 shells is still over-constrained by the data.

To estimate the goodness of fit of a model, we compute the quantity: 
\begin{equation} \sigma = \sum_{filter~i=1}^{filter~i=N} w_{i} \times 
\left[\log (f_{i}^{obs}) - \log (f_{i}^{mod})\right]^2, 
\end{equation}
where $f_{i}^{obs}$ and $f_{i}^{mod}$ are respectively the observed and 
modeled fluxes, the latter being obtained by convolving 
the SED with the appropriate filter bandpasses, and $\rm w_{i}$ is a
weight defined as: 
\begin{equation} \rm w_{i} = \frac{1}{R_{i}^{pow}}
\times \frac{N}{\sum_{filter~i=1}^{filter~i=N} \frac{1}{R_{i}^{pow}}}
\end{equation} 
Here, $\rm R_{i}^{pow}$ is the resolving power
$\lambda/\Delta\lambda$ of each filter or pseudo-filter. 

The weighting scheme is designed 
to reflect the fact that a broadband filter in the NIR collects the same
amount of information as a broadband filter in the FIR, although their
widths are very different. The sum of the weights is normalized
to N, the number of filters used to construct the SED. 

To search for the best fitting dust SED, i.e. the one with the 
lowest $\sigma$, 
we have run about 3$\times$10$^{6}$ \dusty\ models, varying 
$N_{\rm shell}$ from 1 to 3. The optical depth 
$\tau_V$ is varied from 0 to 50, with most of the exploration being done in
the range 0 to 30, with steps of 1 or 2.
The minimum and maximum dust grain sizes are varied within the range between
0.001 and 40\,\micron, with the minimum sizes clustered 
around 0.01\,\micron\
and the maximum sizes clustered around 1\,\micron ; the steps are
generally logarithmic, varying from a 10\% increase to a factor of 2.
The exponent of the size distribution is varied within the range [0-4.5] with
most of the models in the range [2.5-4]. Typical steps are 0.5.
The temperature on the inner side of the dust shell ranged from
10\,K to 1600\,K, with most of the models in the range 400\,K to 700\,K,
and typical steps are 25-50\,K.
Only 3-component dust mixes were investigated (with silicates, graphites
and amorphous carbons), with their proportions varying uniformly  
in the range 0-100\%. Steps in the proportions vary between 1\% to 30\%.
1.6 million 1-zone models were explored, with the outer radius
varying from 25 to 1000 in an uniform manner and with a density
exponent ranging between 0 and 2. 
We ran also just short of a million 2-zone models where the outer
radius was varied between 100 and 1000, and the intermediate radius
between 2 and 999. In these models the density exponent ranged
between $-$2 and 4, with most values between $-$1 and 2.
About 3.5$\times10^5$ 3-zone models were run, 
where the outer radius was fixed at
1000 and the intermediate radius sampled the range from 10 to 999 
uniformly. 
In the 3-zone models, the density exponent was varied from $-1$ to 3 
with a large number of models between 0 and 1.
Steps in radius are between 25 and 50, and steps in the density exponent are 0.5.

\subsection{Results}

\begin{figure}
\hbox{\includegraphics[angle=90,width=1.0\linewidth]{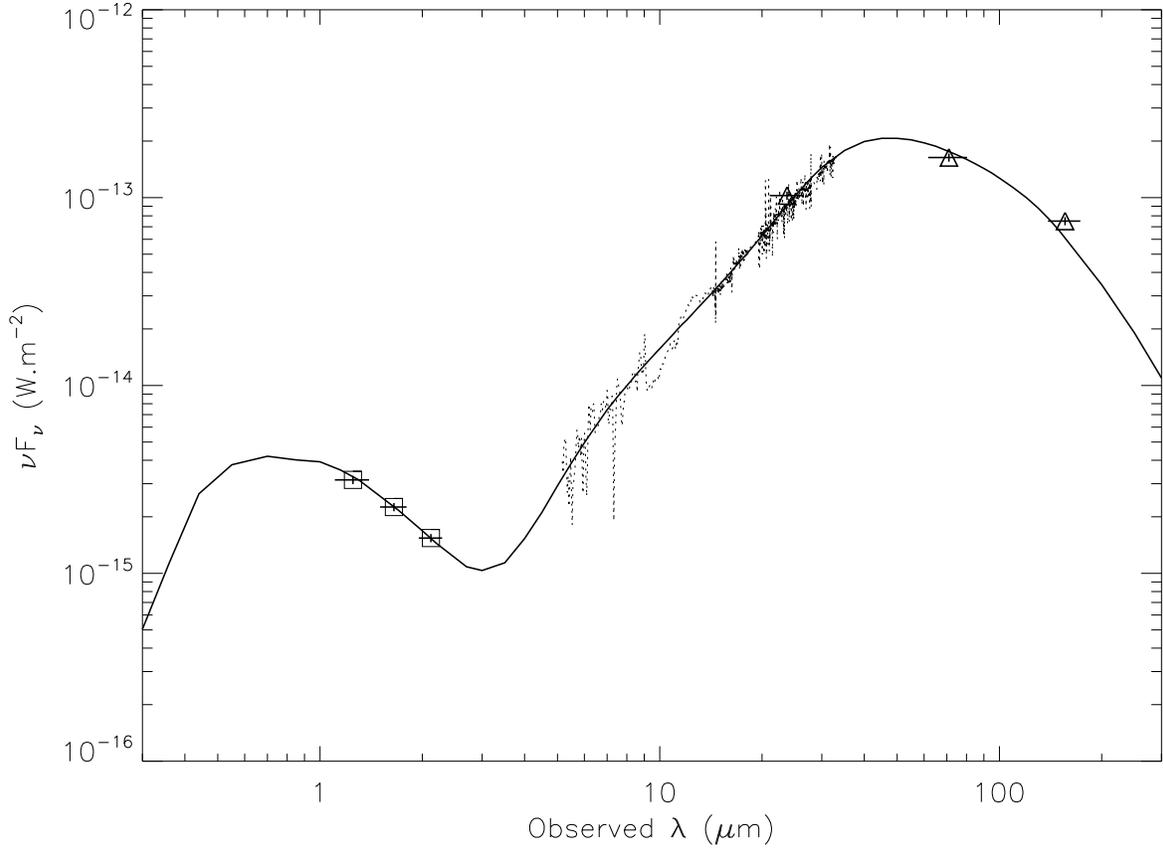} }
\caption{
The \dusty\ fit to the spectral energy distribution of region A
in $\nu F_\nu$ vs. $\lambda$. The
thick continuous line in the model spectrum has been normalized by the median
ratio between the observed and model fluxes. The open squares
represent the NIR photometry while the open triangles represent 
MIPS. For both datasets, the horizontal bar shows the
width of the filter while the vertical bar, generally smaller than
the symbol, is the error bar. The dotted line is the IRS spectrum
from which the PAH features and main ionic lines have been subtracted.
\label{fig:dusty}}
\end{figure}

The parameters of the best-fit \dusty\ model are listed in
Table~\ref{tab:dusty} and the resulting SED is displayed in
Fig.~\ref{fig:dusty}. The $\sigma$ of this model is 0.015, which implies a
mean ratio between the model and observed fluxes of $\sim$7\%,
well within the observational errors. 
Uncertainties in the parameters (last column in Table~\ref{tab:dusty})
are estimated by exploring the range where $\sigma$ is degraded to 10\%.
From the \dusty\ fit, it is
possible to reconstruct physical parameters such as the size of
the system, the bolometric luminosity, and the dust mass, following the 
methods described in \citet{plante02} and \citet{hunt05}.

\subsubsection{The Ionizing Star Cluster}

As always in such a fit, the best constrained quantity is the
bolometric luminosity of the system. This is because, for a
well-sampled SED, this luminosity being the integral of the SED, the
observed and best-fit ones should be the same. The luminosity
of the model is 2.8$\times$10$^{9}$\,\lsun. This is quite a large luminosity
for a single cluster however, as discussed before,
 we are quite likely dealing in this case with at
least two clusters (A1 and A2). The bolometric luminosity constitutes a large
but not unreasonable fraction of the total galaxy luminosity: 
for comparison, the 8-1000\,\mic\ luminosity derived from
the IRAS bands \citep{sanders96} is 4.1$\times$10$^{9}$\,\lsun. With the
bolometric luminosity of the system, we can then use the SB99
models to derive several physical parameters of the central ionizing 
source. The
best-fit SED corresponds to a SB99 model with a Salpeter Initial Mass Function 
and a burst age of 5\,Myr. As this SB99
cluster has a luminosity of 10$^{9}$\,\lsun\ for a stellar
mass of 10$^{6}$\,\msun, the ionizing source has then a mass of
2.8$\times$10$^{6}$\,\msun. This is again in the high mass range for stellar
clusters, but acceptable given that region A is a multiple source. 

The best-fit age for the cluster is 5\,Myr, in the upper range of ages that 
have been derived for region A using various diagnostics.
For example, in the course of a large program  
to determine electron temperatures of \hii\ regions in BCDs by fitting their
optical SEDs (3200--5200 \AA) and in particular  
the Balmer jump ($\lambda$3646 \AA),
\citet{guseva06} have determined the stellar populations of 
a number of galaxies, including region A in \hh. 
They approximate the star formation history of region A by two bursts of 
different strengths, a recent burst with an age t(young) 
to account for the young stellar 
population and a prior burst with an age t(old) 
responsible for the older stars. The contribution 
of each burst to the SED is given by the ratio of the masses of 
stellar populations formed respectively in the old and young bursts,
b= M(old)/M(young). The contribution of gaseous emission is determined 
from the observed equivalent width of the \hb\ emission line EW(\hb). 
Those authors ran a large number (10$^5$) of Monte Carlo models varying 
simultaneously t(young), t(old) and b. 
They found that the model which best fits the SED of \hh\ has t(young) = 
4 Myr, t(old) = 2 Gyr and b = 20. The age of the young burst so derived  
is in good agreement with the one of $\sim$ 5 Myr derived by \citet{johnson04}
from fitting the NIR colors, and the one derived here by the \dusty\ fit.
   
SB99 predicts also the number of hydrogen-ionizing photons \qlyc\  
from such a cluster. We find 
\qlyc\,$\sim\,4.4\times10^{52}$\,photons s$^{-1}$.
This is on the low side of the range of values for A1$+$A2
given by \citet{johnson04}, i.e. from 4.3 (\brg) to 8.2$\times10^{52}$\,photons s$^{-1}$
(radio) converted to the distance used here. 
\citet{johnson04} interpret the disagreement between the two derivations of
\qlyc\ as a sign for the existence of highly obscured sources in region A. 
This explanation cannot account for our low value of \qlyc: since
we reproduce the full SED of region A, there is no room for extra sources as
they would inevitably contribute substantial emission in some wavelength
range. A possible explanation could be that the age we derive is too old. As
can be seen in \citet{sb99} a 2.5 Myr cluster has a \qlyc\ 5 times higher while
its total luminosity has only increased by a factor 1.5. Yet a more likely
explanation is related to the composite nature of region A: A1 seems to suffer
from very little extinction, while A2 seems to be more obscured. One could thus
envision a situation where the less obscured source is indeed a 5 Myr cluster,
though less luminous than we find here, while the more obscured source is a
younger cluster. Such a model however has too many free parameters to be
constrained by the data at hand.

\subsubsection{The Dust Cloud}

The derived properties of the dust ``shell'' around the ionizing clusters 
lead to a more surprising picture. 
The first intriguing feature concerns the nature of the dust.
We find that silicates are absent from the
best-fitting chemical mix. This situation is similar to that
in NGC\,5253 \citep{van04} where 
modeling of the embedded star cluster also indicates 
a lack of silicates. This situation likely results from the 
need for significantly hot dust to produce the NIR continuum, 
while avoiding the generation of silicate emission/absorption features which 
would be expected from small hot silicate grains in an optically thick
regime. If the hot dust continuum were produced by thermally
fluctuating small carbon grains, then colder
silicate dust grains may exist in the outer part envelope. However these two
features -- stochastic heating, and a radial dependence of the chemical
composition -- are not implemented in \dusty.

The second intriguing feature concerns the size of the grains.
To reproduce the full SED, and in
particular its FIR part, the maximum grain size must
exceed $\sim$11\,\micron;
the upper limit to the largest possible size is unconstrained by our fit. 
Even though the best-fit value is large, $\sim$ 40\,\micron,
this is still small compared to the millimeter-sized grains that have been
advocated to fit the SEDs of circumstellar shells \citep[see
e.g.][]{men02}, this is quite large compared to the standard MRN
size distribution \citep{mrn}, with a maximum size of 0.25-1.0\,\micron. If we allow the
fit to the 160\,\mic\ MIPS flux to be significantly worse since it is 
likely an upper limit, we can
relax the constraint on the largest grain size to values that are more
consistent with the MRN distribution. Unlike the maximum size,
the minimum size and the exponent
of the size distribution are very similar to the MRN distribution.

The last surprising feature concerns the extent and  
mass of the ``shell'': at the adopted distance of 16.3\,Mpc, it has a
radius of 430\,pc, considerably larger than region~A and 
nearly reaching region~B (Fig.~\ref{fig:iracoverlay}).
A more detailed look at the model
reveals that, despite the large extent of the dust shell,
the other physical parameters are not unreasonable. 
First, with $\tau_{\rm V} = 3$, the shell needs not be 
very thick. In fact, it is not truly a shell since 
the density shows no dependence with radius, i.e. the exponent of the 
density power law distribution is equal to zero.
Thus, the geometry deduced from the \dusty\ modeling 
is more that of a large cloud of dust surrounding
region~A than that of a giant dust shell. One could in fact consider
that the best fitting model describes, not so much an embedded cluster
and its envelope, but the general ISM of \hh\
heated locally by the clusters in region~A. 
This would make region~A an example of the later stage in the
possible evolutionary sequence that goes from totally embedded
super-star clusters to completely unobscured ones
\citep[e.g. ][]{calzetti97}. 
The IRAC and the shorter-wavelength MIPS images do not show such 
an extended dust cloud  
because of the presence of temperature and hence surface brightness
radial gradients in \hh. In the model, the energy source lies at
the center, the surface brightness is centrally peaked
and the apparent size of the object is smaller than its physical
size. 
At 160\,\micron, with a resolution of $\sim\,40$\arcsec, 
\hh\ appears extended (resolved), which would be consistent
with the cool dust temperature at a radius of 430\,pc.
\dusty\ can compute images at selected wavelengths; 
these images reveal that the FWHM of the dust cloud is always
smaller than the instrumental PSF 24 and 70 \,\micron. 
Therefore although the model dust
cloud is large, it would not necessarily be seen as a resolved object in 
the 24 and 70\,\micron\ images. 
At 160\,\micron\ however, with a resolution of $\sim\,40$\arcsec, 
\hh\ appears slightly resolved, which is not the case of the model cloud. 
This probably indicates that a fraction from the 160\,\micron\ flux is not 
heated by region A but rather finds its energy source in the general 
interstellar radiation field.

We derive an opacity $\tau_{\rm V} = 3$ from our best-fit model.
In the model context, this is the
optical depth of the ``shell'' toward the central
source. There are many possible explanations of the difference
between the \dusty -derived optical depth $\tau$ and the measured values of the
extinction $A$ from recombination lines. The simple and linear relation between the two
quantities ($A \sim 1.086\,\tau$) is only valid in the screen
geometry, when scattering and radiation transfer are neglected.
Although \dusty\ assumes a screen geometry, it
treats the complete radiation transfer process: absorption, scattering, 
and re-emission of the absorbed energy. 
As scattering has a wavelength dependence opposite to that of extinction, 
the resulting relation
between $A$ and $\tau$ becomes shallower and depends strongly on 
wavelength. Another plausible explanation is that a more realistic geometry would have
part of the source region, and in particular the ionized gas
responsible for all the lines and for a large fraction of the NIR
emission, mixed with the absorbing dust. 
This also results in a shallower relation between $A$ and $\tau$. 
In any case,
the model value of \av\ is within the range of observationally determined values 
of \av. From the optical recombination lines,
\citet{thuanizotov05} derive \av\,=\,0.5. 
Comparing Br$\gamma$ from \citet{johnson04} with \hb\ gives \av\,=\,0.9.
From the location of
region~A in the (J-H/H-K) color-color diagram, \citet{johnson04}
estimate a larger \av\ in the range 2-4. Finally, the comparison of
\qlyc\ from the radio emission to that deduced from
the \brg\ line gives a still larger value, \av\,=\,8
\citep{johnson04}. There thus appears to be a systematic increase of the 
derived extinction with increasing wavelengths.
These differences may be due to a variety of causes such as: 
(1)~mixing between
the emission and absorption regions; (2)~the coexistence of sources with
different extinction within the Spitzer beam; or (3)~the clumpy
distribution of the absorbing material. For lack of a better spatial
resolution in the infrared, it is impossible to distinguish among these
possibilities, although we do know that region~A
comprises three components in the radio.
That we can model the complete NIR-FIR SED
with a characteristic opacity in the intermediate range of the 
observationally determined 
extinctions suggests that our model is representative of the
{\it average} situation in region~A.

From the dust properties and the physical size of the system, we can
compute the total dust mass which we find to be 
1.5$\times10^{6}$\,\msun. This is again a large dust mass for 
embedded clusters, which generally are one order of 
magnitude less massive. It is however a quite reasonable mass for a dust cloud 
in the general ISM of \hh. Indeed, based on the observations of 
\citet{meier01} and \citet{gordon81}, \citet{johnson04} have derived a
total molecular and atomic gas mass of $\sim$\,6.7$\times10^{8}$\,\msun.
Thus \hh\ has a dust-to-gas ratio of 2.2$\times10^{-3}$, consistent
with the metallicity trends for BCDs illustrated by \citet{hunt05}
and references therein.

To summarize, a model where the stellar clusters
in region~A heat an extended and diffuse region of the ISM of \hh\ 
can account well for the NIR-FIR spectral energy distribution of
the galaxy. 

\section{Star Formation in \hh\ \label{sec:discussion} }

Star formation in \hh\ occurs primarily in several knots distributed
over the central region of the galaxy around the dust lane.
The primary star-forming regions,
knots A, B, and C (see Fig. \ref{fig:irac}) show different
properties in the optical, the infrared, and the radio.
Because of the compact radio emission found by \citet{johnson04}, we
would like to argue that knot A is a ensemble of star clusters
with generally large extinction (the extinction is highly variable),
while knot B is older and less obscured.
Knot C, with its seemingly high PAH fraction, would be older still,
with an \ha\ flux $\sim$6 times smaller than the other two knots,
but with similar extinction \citep{steel96}.
Because of its high density 
\citep[n$_{\rm e}\sim$200\,cm$^{-3}$][]{steel96,izotovthuan04} 
and compact size in the radio \citep{johnson04},
we would classify knot A as ``active''
\citep[see the definition of this term in][]{hirashita04}.
Such conditions are found also in mergers, with consequently a high
pressure, and possible formation of super star clusters.
That knot A is forming stars in an ``active'' mode is corroborated by 
the good fit of the \dusty\ model to the IR SED, the significant IR luminosity,
and the relatively high SFR inferred by \citet{johnson04} from the
radio emission.
Knots B and C, on the other hand, are less dense
\citep[n$_{\rm e}\sim$100\,cm$^{-3}$][]{steel96}, and can be considered as
''passive''. 

The properties of the knots differ in terms of their IRAC and MIPS colors,
but the IRS spectrum encompasses variable fractions of knots A and B,
as well as the intervening ISM in the dust lane dividing
the two regions. 
Were knot A to be an isolated dusty star cluster with a rising-spectrum
radio source, we would expect it to have little or no PAH emission
\citep{hunt05} because
we would expect the PAHs to be destroyed by the intense radiation field in
the ``active'' compact region.
However, we do see PAH emission.
Therefore either our hypothesis is mistaken, namely PAHs can be
found in compact dense star-forming regions, or our spectrum is 
``contaminated'' by region~B and the global ISM in \hh\ (or both).
To distinguish among the various possibilities,
more extensive samples are needed, with a wide range
of metallicities, luminosities, and PAH properties.

\hh\ is relatively nearby BCD, only slightly
metal deficient, with 12$+$log(O/H)\,=\,8.32, and only slightly
sub-luminous, with $M_{\rm B}=-17.9$. 
Hence, it is an object which could serve as a transition case
to link the properties of low-metallicity BCDs with those of
more luminous metal-rich starbursts.
Indeed, the EWs of its PAHs are generally higher than those of
other BCDs \citep[e.g. ][]{wu06}, more consistent
with more luminous and metal-rich starburst galaxies \citep{brandl04}.
On the other hand, the PAHs in \hh\ tend to be neutral according to 
the diagnostic diagram of \citet{draine01}, more similar to ``normal''
galaxies than to starbursts.
PAHs in BCDs may also be narrower than those of starbursts, and more similar
to nearby \hii\ regions although the differences may not be significant.
The problem with many of these results is that their exact interpretation
depends on the fraction of the global ISM in the IRS slits.
These appear to cross the dust lane that separates the apparent positions
of star-forming knots A and B, so
we could be seeing a PAH contribution from the general ISM, not 
necessarily associated with the star formation.
The PAH parameters also depend on how they were derived 
(e.g., continuum determination, profile templates, etc.).
Consequently, on the basis of a single object, we cannot
make definitive statements about low-metallicity PAHs,
and must wait for larger BCD
and starburst samples analyzed in a homogeneous way.


\section{Summary\label{sec:conclusions} }

\begin{figure}
\vspace*{-15mm}
\vbox{
\includegraphics[angle=0,width=1.0\linewidth,bb=18 165 590 548]{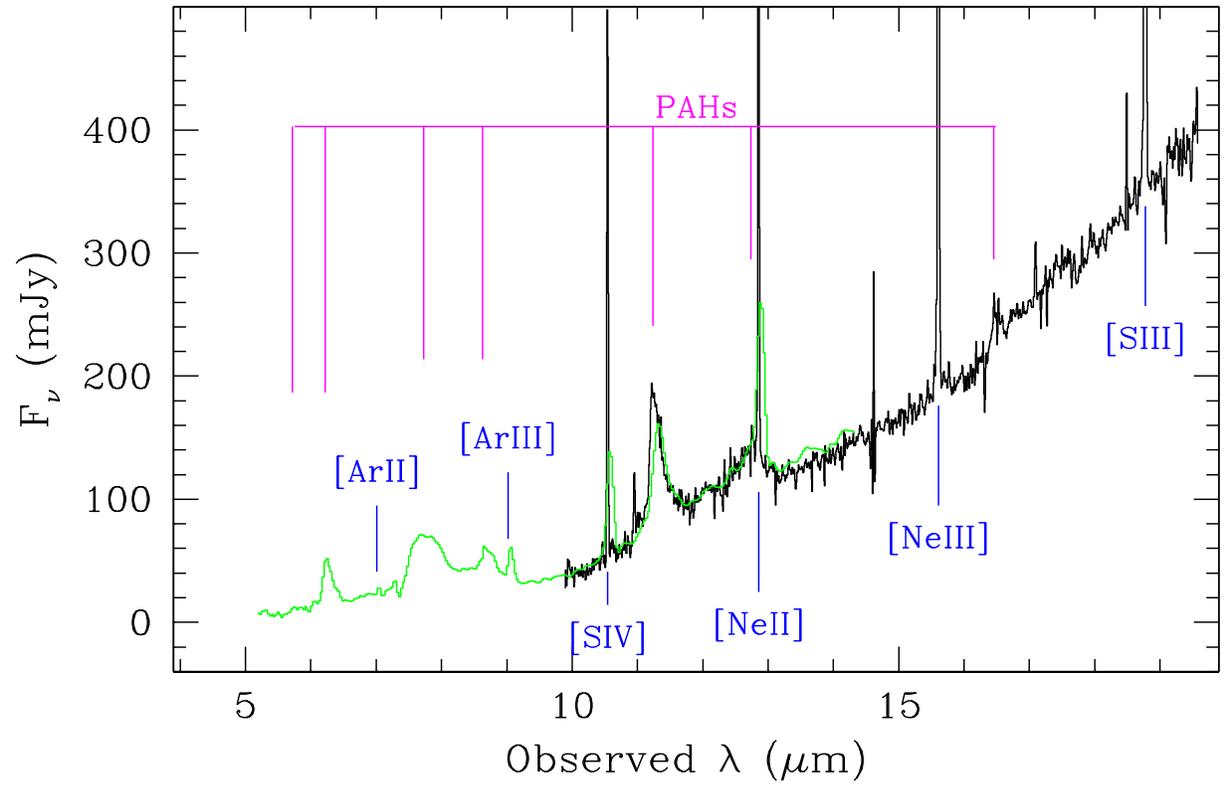}
\includegraphics[angle=0,width=1.0\linewidth,bb=18 165 590 548]{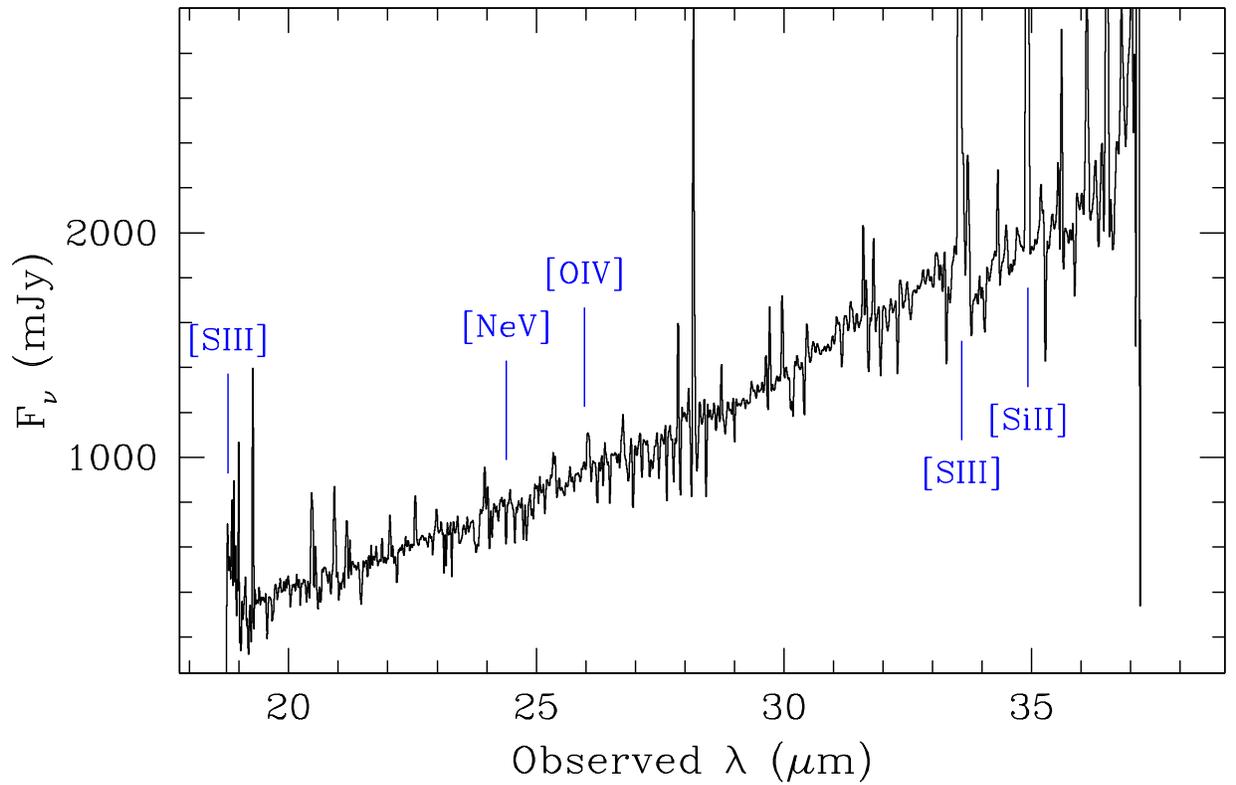}
}
\caption{The IRS spectrum of \hh\ consisting of the short-wavelength low-resolution mode
and both high-resolution modes:
SL$+$SH are shown in the upper panel, and $LH$ in the lower one. 
The PAH features at 5.7, 6.2, 7.7, 8.6, 11.2, 12.7, and 16.4 \micron\
are clearly detected, as are several emission lines:
\siv, \neii, \neiii, \siii, \oiv.
\label{fig:irs}}
\end{figure}

We present here the first results of our 
\spitzer\ Cycle 1 program which is aimed at the investigation of star
formation in metal-poor BCD galaxies. We wish to 
study how star formation parameters 
and dust properties change as a function of metallicity and excitation. 
In this first paper of a series, we discuss  
the infrared properties of the most metal-rich object in our sample, 
\hh\ = NGC 3353 = Mrk 35, with an oxygen abundance of 12$+$log(O/H)\,=\,8.32. 
We have obtained IRS low and high-resolution 
spectra for the brightest star-forming region A in \hh, as well as 
IRAC images at 4.5 and 8\,\micron\ and MIPS images at 
24, 70, and 160\,\micron\ of the galaxy. We obtain the following results:

\noindent
(1) The IRS spectrum (Fig. \ref{fig:irs}) 
shows strong Polycyclic Aromatic Hydrocarbon (PAH) molecular  
emission, with features clearly detected at 5.7, 6.2, 7.7, 8.6, 11.2, 
12.7, and 16.4\,\micron. The PAHs in \hh\ are predominantly neutral and 
small, similar to those found in normal spiral galaxies, suggesting that 
they reside in the general ISM and not in the star-forming region A. The PAH 
emission features are relatively narrow and their equivalent widths 
are generally high for the metallicity of \hh.
   
\noindent
(2) Gaseous nebular line emission is also seen.
The IRS spectrum shows several fine-structure lines, including
\siv $\lambda$10.51, \neii $\lambda$12.81,       
\neiii $\lambda$15.55, 36.01, \siii $\lambda$18.71, 33.58,
\oiv $\lambda$25.89, and 
\SiII $\lambda$34.82\,\micron. 

\noindent
(3) Several fairly high-excitation lines are detected in \hh.
The faint \oiv\ line at 25.89\,\micron\ indicates 
the presence of radiation as hard as 54.9 eV in \hh. This hard radiation 
is perhaps due to the Wolf-Rayet stars present in the galaxy. 
Such a detection is intriguing because the high-ionization \fev $\lambda$4227 
and \heii\ $\lambda$4686 emission lines which both have an 
ionization potential of 54.4\,eV, are not seen in the optical spectrum of \hh,
presumably because of extinction effects.
The electron density derived from the MIR lines is about half 
that derived from the optical lines. Moreover 
\hii\ photoionization CLOUDY models that reproduce the optical 
line intensities predict MIR lines that are systematically too small 
compared to those observed.  
This discrepancy may be understood if the  
beam dilution in the IR is taken into account, and if the bulk of the MIR 
radiation comes from a region more obscured than the optical one.

\noindent
(4) Like most galaxies, \hh\ is a composite entity in the IR.
We see {\it extended} stellar photospheric emission
and {\it compact} hot dust continuum at 4.5\,\micron\ coming mainly from the 
star-forming regions, {\it extended} PAH emission coming mainly from 
the general ISM at 8\,\micron, with a small  
contribution from hot dust continuum and stars, 
{\it compact} small grain warm dust emission associated with 
active star-forming regions at 24 and 70\,\micron, 
and cooler {\it extended} dust 
emission associated with older stellar populations at 160\,\micron.  

\noindent
(5) We have modeled the IR spectral energy distribution (SED) of star-forming 
region A with \dusty, a code which solves the radiative transfer problem in a 
dust envelope. The best-fit model gives a total luminosity of 
2.8$\times$10$^9$ L$_\sun$ and a total stellar mass of 2.8$\times$10$^6$ M$_\sun$ for 
the ionizing star clusters.
These numbers are not unreasonably high because region A contains at least three 
star clusters. The best-fit age of the star formation burst in A is 5\,Myr, in 
agreement with the age derived from the optical SED. The number 
of ionizing photons is 4.4$\times10^{52}$ photons s$^{-1}$, greater 
than the one derived from the Br$\gamma$ line but smaller than that
derived from the thermal radio flux, indicating more extinction in  
the Br$\gamma$ flux than in the MIR. 
 
\noindent
(6) As for the nature of the dust, the best-fit \dusty\ model implies that 
silicates are absent from region A. The maximum grain size can reach the very 
large value of 40 \micron\ if we attempt to fit the 160 \micron\ MIPS flux.
However, we obtain more reasonable results
($\sim$ 1 \micron) if we consider the 160 \micron\ 
flux to be an upper limit and do not attempt to fit it. The 
\av\ of the dust given by \dusty\ is $\la$ 3, higher than the 
value of 0.5 derived from the optical lines, and lower than 
the value of $\sim$ 8 derived from radio observations, consistent 
with the picture that observations at longer wavelengths probe 
more deeply into the star-forming region. The dust 
density does not vary with distance from the central clusters, and   
the dust extent is 430\,pc in radius, considerably larger than region~A. This 
suggests that the geometry deduced from the \dusty\ modeling 
is that of a large cloud of dust in the general ISM of \hh, surrounding
region~A and heated by the clusters in it, rather
than that of a giant dust shell around region~A. 

\acknowledgments

This work is based on observations made with the Spitzer
Space Telescope, which is operated by the Jet Propulsion Laboratory, 
California Institute of Technology, under NASA contract 1407. 
We would like to thank Yanling Wu and Jim Houck for astute comments,
and an anomymous referee for a careful reading and insightful suggestions
which improved the paper.
Support for this work was provided by NASA through contract GG10638 
issued by JPL/Caltech. T.X.T. and Y.I.I. acknowledge partial financial 
support from NSF grant AST 02-05785. T.X.T. is grateful for the hospitality of 
the Service d'Astrophysique at Saclay and of the Institut d'Astrophysique 
in Paris during his sabbatical leave. He thanks the University of Virginia 
for the award of a Sesquicentennial Fellowship. 
This research has made extensive use of the NASA/IPAC Extragalactic Database (NED), operated 
by the Jet Propulsion Laboratory, California Institute of Technology, under 
contract with the National Aeronautics and Space Administration.

\clearpage

\begin{deluxetable}{ccc}
\tablecaption{IRAC and MIPS total fluxes of Haro\,3 with Other 
Published Photometry \label{tab:photom} }
\tablewidth{0pt}
\tablehead{
\colhead{Telescope/Instrument} & \colhead{Wavelength} & \colhead
{Total flux} \\
\colhead{} & \colhead{(\mic)} & \colhead{(mJy)} \\
}
\startdata
Spitzer/IRAC & 4.509 & 20.7 $\pm$ 2.1  \\
" & 7.982 & 148.0 $\pm$ 14.8 \\
Spitzer/MIPS & 23.7 & 810 $\pm$ 81 \\
" & 71.0 & 3870 $\pm$ 387 \\
" & 156.0 & 3900 $\pm$390 \\
\hline
IRAS\tablenotemark{a} & 12 & 211.1 $\pm$ 29.6 \\
"                     & 25 & 941.8 $\pm$ 56.5 \\
"                     & 60 & 4949.0 $\pm$ 395.9 \\
"                     & 100 & 6746.0 $\pm$ 404.8 \\
\hline
ISO/ISOCAM\tablenotemark{b} & 7.75 & 149 $\pm$ 30 \\
"                           & 15.0 & 300 $\pm$ 70 \\
\hline
WIYN/NIRIM\tablenotemark{c} & 1.26 & 1.19 $\pm$ 0.12 \\
"                       & 1.65 & 1.12 $\pm$ 0.11 \\
"                       & 2.12 & 1.06 $\pm$ 0.11 \\
\hline
VLA\tablenotemark{c} & 13000 & 2.96 $\pm$ 0.22 \\
"                    & 36000 & 2.60 $\pm$ 0.22 \\
\enddata
\tablenotetext{a}{From NED.}
\tablenotetext{b}{From \citet{metcalfe96}.}
\tablenotetext{c}{From \citet{johnson04} for A1+A2.}
\end{deluxetable}

\begin{deluxetable}{rccl}
\footnotesize
\tablecaption{PAH Parameters Obtained by Fitting Lorentzian Profiles\label{tab:pahs}}
\tablewidth{0pt}
\tablehead{
\multicolumn{1}{c}{Wavelength\tablenotemark{a}} &
\multicolumn{1}{c}{Integrated\tablenotemark{b}} &
\multicolumn{1}{c}{Equivalent\tablenotemark{b}} &
\multicolumn{1}{c}{FWHM\tablenotemark{b}} \\
& \multicolumn{1}{c}{Flux} &
\multicolumn{1}{c}{Width} \\
\colhead{(\micron)}&\colhead{($10^{-16}$\,W\,m$^{-2}$)}
&\colhead{(\micron)}&\colhead{(\micron)}}
\startdata
6.202 	&  ~~8.3 (0.3) & 1.31 (0.22) & 0.17 (0.004) \\
7.693 	&  26.0 (9.2) & 2.85 (1.47) & 0.66 (0.11)  \\
8.631   &  ~~6.4 (0.9) & 0.64 (0.12) & 0.38 (0.08)  \\
11.269	&  ~~8.3 (0.2) & 0.60 (0.05) & 0.25 (0.006) \\
12.814	&  ~~6.3 (1.3) & 0.32 (0.07) & 0.13 (0.015) \\
16.424	&  ~~1.2 (0.1) & 0.06 (0.01) & 0.18 (0.007) \\
\enddata
\tablenotetext{a}{Fitted restframe wavelength, corrected 
for $z=0.00323$.}
\tablenotetext{b}{Standard deviations of the repeated measurements
 are given in parentheses.
The true uncertainty including calibration is probably $\sim$20\%.}
\end{deluxetable}    

\begin{deluxetable}{crccccc}
\footnotesize
\tablecaption{Fine Structure Line Parameters 
Obtained by Fitting Gaussian Profiles\label{tab:lines}}
\tablewidth{0pt}
\tablehead{
\multicolumn{1}{c}{Line\tablenotemark{a}} &
\multicolumn{1}{c}{E$_{\rm ion}$\tablenotemark{b}} &
\multicolumn{1}{c}{Nominal} &
\multicolumn{1}{c}{Fitted\tablenotemark{c}} &
\multicolumn{1}{c}{Integrated\tablenotemark{d}} &
\multicolumn{1}{c}{Equivalent\tablenotemark{d}} &
\multicolumn{1}{c}{FWHM\tablenotemark{d}} \\
&& \multicolumn{1}{c}{Wavelength} &\multicolumn{1}{c}{Wavelength} 
& \multicolumn{1}{c}{Flux} & \multicolumn{1}{c}{Width} \\
&\colhead{(eV)} & \colhead{(\micron)} & \colhead{(\micron)} & 
\colhead{($10^{-16}$\,W\,m$^{-2}$)}
&\colhead{(\micron)} & \colhead{(\micron)}}
\startdata
\siv   & 34.8 & 10.511  & 10.510 &  ~~3.01 (0.20) & 0.21 (0.01) & 0.022 (0.005) \\
\neii  & 21.6 & 12.814 	& 12.813 &  ~~2.71 (0.05) & 0.12 (0.01) & 0.020 (0.001) \\
\neiii & 41.0 & 15.555	& 15.556 &  ~~7.51 (0.07) & 0.31 (0.02) & 0.022 (0.001) \\
\siii  & 23.3 & 18.713	& 18.713 &  ~~4.03 (0.01) & 0.13 (0.01) & 0.030 (0.001) \\
\oiv   & 54.9 & 25.890	& 25.961 &  ~~0.60 (0.02) & 0.02 (0.01) & 0.065 (0.002) \\
\siii  & 23.3 & 33.481	& 33.439 &   10.6~~ (0.01)  & 0.22 (0.01) & 0.055 (0.001) \\
\SiII  & ~~8.2  & 34.815  & 34.800 &  ~~4.80 (0.01) & 0.10 (0.01) & 0.056 (0.001) \\
\neiii & 41.0 & 36.014 	& 36.000 &  ~~1.27 (0.07) & 0.31 (0.02) & 0.66 (0.11)  \\
\enddata
\tablenotetext{a}{Because of the presence of \ariii\ and \arii\ only in the low-resolution
spectra, we do not give their parameters here.}
\tablenotetext{b}{Ionization potential (lower) of the stage leading to the transition.}
\tablenotetext{c}{Restframe, corrected for the $z=0.00323$
\citep{thuanizotov05}.}
\tablenotetext{d}{Standard deviations of the repeated measurements
 are given in parentheses.
The true uncertainty including calibration is probably $\sim$20\%.}
\end{deluxetable}

\begin{deluxetable}{lc}
\tablecaption{Chemical Composition of Regions A1 and A2 used in CLOUDY
Calculations \label{tab:chem_cloudy}}
\tablewidth{0pt}
\tablehead{\colhead{Element} & \colhead{log X/H}
}
\startdata
He      &        $-$1.07 \\
C       &        $-$4.03 \\
N       &        $-$4.93 \\
O       &        $-$3.68 \\
Ne      &        $-$4.53 \\
Si      &        $-$5.13 \\
S       &        $-$5.48 \\
Cl      &        $-$7.03 \\
Ar      &        $-$6.08 \\
Fe      &        $-$5.78 \\
\enddata
\end{deluxetable}

\begin{deluxetable}{lcc}
\tablecaption{Comparison of Observed and Modeled Optical Relative Line
Intensities in Region A1 \label{tab:optical_cloudy}}
\tablewidth{0pt}
\tablehead{\colhead{Line} & \colhead{Observations}\tablenotemark{a} & \colhead{CLOUDY}\tablenotemark{a}
}
\startdata
3727 [O {\sc ii}]       &2.4978   &2.3101 \\
3750 H12                &0.0374   &0.0314 \\
3771 H11                &0.0433   &0.0407 \\
3798 H10                &0.0553   &0.0542 \\
3835 H9                 &0.0761   &0.0746 \\
3869 [Ne {\sc iii}]     &0.3141   &0.3191 \\
4068 [S {\sc ii}]       &0.0215   &0.0225 \\
4101 H$\delta$          &0.2817   &0.2612 \\
4340 H$\gamma$          &0.4966   &0.4705 \\
4363 [O {\sc iii}]      &0.0274   &0.0272 \\
4658 [Fe {\sc iii}]     &0.0091   &0.0094 \\
4861 H$\beta$           &1.0000   &1.0000 \\
4959 [O {\sc iii}]      &1.3176   &1.3660 \\
5007 [O {\sc iii}]      &4.0352   &4.1118 \\
5200 [N {\sc i}]        &0.0072   &0.0052 \\
5270 [Fe {\sc iii}]     &0.0047   &0.0054 \\
5518 [Cl {\sc iii}]     &0.0059   &0.0061 \\
5538 [Cl {\sc iii}]     &0.0042   &0.0044 \\
5755 [N {\sc ii}]       &0.0037   &0.0039 \\
6300 [O {\sc i}]        &0.0094   &0.0504 \\
6312 [S {\sc iii}]      &0.0123   &0.0120 \\
6563 H$\alpha$          &2.9491   &2.8930 \\
6583 [N {\sc ii}]       &0.2007   &0.2025 \\
6717 [S {\sc ii}]       &0.1571   &0.1991 \\
6731 [S {\sc ii}]       &0.1267   &0.1506 \\
7135 [Ar {\sc iii}]     &0.0929   &0.0888 \\
7320 [O {\sc ii}]       &0.0259   &0.0288 \\
7330 [O {\sc ii}]       &0.0223   &0.0234 \\
\enddata
\tablenotetext{a}{Relative to H$\beta$.}
\end{deluxetable}    

\begin{deluxetable}{lccccc}
\tablecaption{Comparison of the Observed and Predicted Relative
Intensities of the MIR lines \label{tab:mir_cloudy}}
\tablewidth{0pt}
\tablehead{\colhead{Line} & \colhead{Observations} & 
\multicolumn{3}{c}{Model} & \colhead{Obs./Model} \\ \cline{3-5}
\colhead{} & \colhead{A1+A2\tablenotemark{a}} & \colhead{A1\tablenotemark{b}} 
& \colhead{A2\tablenotemark{c}} & \colhead{A1+A2\tablenotemark{a}}
& \colhead{}
}
\startdata
10.51 \siv   &0.26   &0.15 &0.08 &0.11   &2.4 \\
12.81 \neii  &0.24   &0.03 &0.05 &0.04   &6.0 \\
15.55 \neiii &0.66   &0.40 &0.37 &0.38   &1.7 \\
18.71 \siii  &0.35   &0.22 &0.21 &0.21   &1.7 \\
33.48 \siii  &0.92   &0.43 &0.41 &0.42   &2.2 \\
34.82 \SiII  &0.42   &0.20 &0.28 &0.25   &1.6 \\
36.01 \neiii &0.11   &0.04 &0.03 &0.03   &3.7 \\
\enddata
\tablenotetext{a}{Relative to the total intensity of the H$\beta$ emission
line in regions A1+A2.}
\tablenotetext{b}{Relative to the intensity of the H$\beta$ emission
line in region A1.}
\tablenotetext{c}{Relative to the intensity of the H$\beta$ emission
line in region A2.}
\end{deluxetable}

\begin{deluxetable}{lccc}
\tablecaption{Results from the \dusty\ Fit \label{tab:dusty} }
\tablewidth{0pt}
\tablehead{
\colhead{Parameter Name} & \colhead{Unit} & \colhead{Best fit value} & \colhead{Confidence range} \\
\colhead{(1)\tablenotemark{a}} & \colhead{(2)\tablenotemark{b}} & 
\colhead{(3)\tablenotemark{c}} & \colhead{(4)\tablenotemark{d}}
}
\startdata
Starburst age & Myr & 5 &  $-$\\
Silicates abundance & \% & 0 & 0$-$5\\
Graphite abundance & \% & 67 & 55$-$77\\
Amorphous Carbon abundance & \% & 33 & 23$-$45\\
a(min) & \mic & 0.008 & 0.006$-$0.010\\
a(max) & \mic & 40.0 & $>$10.8\\
Size distribution exponent & n.a. & 4 & $-$\\
R$^{\rm out}$ & pc & 435 & 380$-$490\\
Density exponent in shell & n.a. & 0 & $-$\\
T$_{\rm int}$ & K & 500 & 479$-$523\\
$\tau_{\rm V}$ & & 3 & 2.8$-$3.3\\
\enddata
\tablenotetext{a}{Parameter name.}
\tablenotetext{b}{Physical units of this parameter.}
\tablenotetext{c}{Best-fit value (mean error per pseudo-filter band is 7\%).}
\tablenotetext{c}{Range within which the change of this parameter only will 
result in a mean error per pseudo-filter band of 10\%.
A $-$ corresponds to values which are so tightly constrained by our fit
that any change in them results in a mean error of $>$10\%.}
\end{deluxetable}

\clearpage

\clearpage
\thispagestyle{empty}
\clearpage
\clearpage
\thispagestyle{empty}

\clearpage

\end{document}